\documentclass[12pt,prd,showpacs,tightenlines,nofootinbib]{revtex4}
\usepackage{bm}
\usepackage{graphics}
\usepackage{rotating}
\usepackage{epsfig}
\begin{document}
\title{\begin{flushright}{\rm\normalsize HU-EP-03/45\\
hep-ph/0308149}\end{flushright}
Weak decays of the $\bm{B_c}$ meson to $\bm{B_s}$ and $\bm{B}$ mesons
in the relativistic quark model}  
\author{D. Ebert}
\affiliation{Institut f\"ur Physik, Humboldt--Universit\"at zu Berlin,
Newtonstr. 15, D-12489  Berlin, Germany}
\author{R. N. Faustov}
\author{V. O. Galkin}
\affiliation{Institut f\"ur Physik, Humboldt--Universit\"at zu Berlin,
Newtonstr. 15, D-12489 Berlin, Germany}
\affiliation{Russian Academy of Sciences, Scientific Council for
Cybernetics, Vavilov Street 40, Moscow 117333, Russia}
\begin{abstract}
Semileptonic and nonleptonic decays of the $B_c$ meson to $B_s$
and $B$ mesons, caused by the $c\to s,d$ quark transitions, are studied
in the framework of the relativistic quark 
model. The heavy quark expansion in inverse powers of the active $c$
and spectator $\bar b$ quark is used to simplify calculations while
the final $s$ and $d$ quarks in the $B_s$ and $B$ mesons are
treated relativistically. The decay form factors are explicitly 
expressed through the overlap integrals of the meson wave functions
in the whole accessible kinematical range. The
obtained  results are compared with the predictions of other
approaches.  
\end{abstract}

\maketitle

\section{Introduction}
\label{sec:intro}

The $B_c$ meson discovered at Fermilab \cite{cdfcol} is the only
quark-antiquark bound system ($\bar b c$) composed of heavy quarks
($b$, $c$) with different flavours, thus being flavour asymmetric. The
investigation of the $B_c$ meson properties (mass spectrum, decay
rates, etc.) is therefore of special interest compared to symmetric
heavy quarkonium ($b\bar b$, $c\bar c$) ones. The difference of quark
flavours forbids the 
annihilation of $B_c$ into gluons. As a result the pseudoscalar $\bar
bc$ state is much more stable than the heavy quarkonium one and decays
only weakly. It serves as a final state for the pionic and radiative
decays of the excited $\bar b c$ states (lying below the $BD$
threshold). Experimental study of the $B_c$ mesons is planned both at
the Tevatron and  Large Hadron Collider (LHC) (for a recent review see
e.g. \cite{gklry} and references therein).

Since both quarks ($b$, $c$) composing the $B_c$ meson are heavy,
their weak decays contribute comparably to the total decay rate. Thus
there are two distinctive decay modes: ($i$) $\bar b\to \bar c, \bar
u$ with $c$ quark being a spectator, and ($ii$) $c\to s,d$ with $\bar b$
quark being a spectator. The transition ($i$) induce the semileptonic
$B_c$ decays to charmonium and $D$ mesons, while the transitions
($ii$) cause the $B_c$ decays to $B_s$ and $B$ mesons. The kinematical
ranges of these semileptonic decay modes are substantially
different. The square of the four momentum transfer to the lepton pair
extends from 0 to  $q^2_{\rm max}\approx 10$~GeV$^2$ for the decays to
charmonium and $q^2_{\rm max}\approx 18$~GeV$^2$ for decays to $D$
mesons, but only to  $q^2_{\rm max}\approx 1$~GeV$^2$ for decays to
$B$ and $B_s$ mesons. Thus the kinematical range for the decay mode
($i$) is appreciably larger than for the decay mode ($ii$). Otherwise
in the $B_c$ rest frame the maximum recoil three momentum of the final
charmonium and $D$ meson turns out to be of order of their masses,
while that of final $B$ and $B_s$ mesons is much smaller than the
meson masses. 

The weak $B_c$ decays to charmonium and $D$ mesons were studied at
length in our recent paper \cite{bcbd}. Here we consider the weak
$B_c$ decays to $B_s$ and $B$ mesons within the relativistic quark
model. The model is based on the quasipotential approach in quantum
field theory and was fruitfully applied for describing the electroweak
decays and mass spectra of heavy-light mesons, heavy quarkonia
\cite{mass1,mass,hlm,gf,fg,mod} and $B_c$ meson \cite{efgbc}. The
relativistic wave functions obtained in the latter paper are used
below to calculate the transition matrix elements. The consistent
theoretical description of $B_c$ decays requires a reliable
determination of the $q^2$ dependence of the decay amplitudes in the
whole kinematical range. In most previous calculations the
corresponding decay form factors were determined only at one
kinematical point either $q^2=0$ or $q^2=q^2_{\rm max}$ and then
extrapolated to the allowed kinematical range using some
phenomenological ansatz (mainly (di)pole or Gaussian). Our aim is to
explicitly  determine the $q^2$ dependence of form factors in the
whole kinematical range thus avoiding extrapolations and reducing
uncertainties.

The paper is organized as follows. In Sec.~\ref{rqm} we describe the
underlying 
relativistic quark model. The method for calculating matrix elements
of the weak current for $c\to s,d$ transitions in $B_c$ meson decays
is presented in Sec.~\ref{mml}. Special attention is paid to the
dependence on the momentum transfer of the decay amplitudes. The $B_c$
decay form factors are calculated in the whole kinematical range in
Sec.~\ref{dff}. The $q^2$ dependence of the form factors is explicitly
determined. These form factors are used for the calculation of the
$B_c$ semileptonic decay rates in Sec.~\ref{ssd}. Section~\ref{nl}
contains our predictions for the energetic nonleptonic $B_c$ decays in the
factorization approximation, and a comparison of our results with other
theoretical calculations is presented. Our conclusions are given in
Sec.~\ref{sec:conc}. Finally, the Appendix contains complete 
expressions for the decay form factors.

\section{Relativistic quark model}  
\label{rqm}

In the quasipotential approach a meson is described by the wave
function of the bound quark-antiquark state, which satisfies the
quasipotential equation \cite{3} of the Schr\"odinger type \cite{4}
\begin{equation}
\label{quas}
{\left(\frac{b^2(M)}{2\mu_{R}}-\frac{{\bf
p}^2}{2\mu_{R}}\right)\Psi_{M}({\bf p})} =\int\frac{d^3 q}{(2\pi)^3}
 V({\bf p,q};M)\Psi_{M}({\bf q}),
\end{equation}
where the relativistic reduced mass is
\begin{equation}
\mu_{R}=\frac{E_1E_2}{E_1+E_2}=\frac{M^4-(m^2_1-m^2_2)^2}{4M^3},
\end{equation}
and $E_1$, $E_2$ are the center of mass energies on mass shell given by
\begin{equation}
\label{ee}
E_1=\frac{M^2-m_2^2+m_1^2}{2M}, \quad E_2=\frac{M^2-m_1^2+m_2^2}{2M}.
\end{equation}
Here $M=E_1+E_2$ is the meson mass, $m_{1,2}$ are the quark masses,
and ${\bf p}$ is their relative momentum.  
In the center of mass system the relative momentum squared on mass shell 
reads
\begin{equation}
{b^2(M) }
=\frac{[M^2-(m_1+m_2)^2][M^2-(m_1-m_2)^2]}{4M^2}.
\end{equation}

The kernel 
$V({\bf p,q};M)$ in Eq.~(\ref{quas}) is the quasipotential operator of
the quark-antiquark interaction. It is constructed with the help of the
off-mass-shell scattering amplitude, projected onto the positive
energy states. 
Constructing the quasipotential of the quark-antiquark interaction, 
we have assumed that the effective
interaction is the sum of the usual one-gluon exchange term with the mixture
of long-range vector and scalar linear confining potentials, where
the vector confining potential
contains the Pauli interaction. The quasipotential is then defined by
\cite{mass1}
  \begin{equation}
\label{qpot}
V({\bf p,q};M)=\bar{u}_1(p)\bar{u}_2(-p){\mathcal V}({\bf p}, {\bf
q};M)u_1(q)u_2(-q),
\end{equation}
with
$${\mathcal V}({\bf p},{\bf q};M)=\frac{4}{3}\alpha_sD_{ \mu\nu}({\bf
k})\gamma_1^{\mu}\gamma_2^{\nu}
+V^V_{\rm conf}({\bf k})\Gamma_1^{\mu}
\Gamma_{2;\mu}+V^S_{\rm conf}({\bf k}),$$
where $\alpha_s$ is the QCD coupling constant, $D_{\mu\nu}$ is the
gluon propagator in the Coulomb gauge
\begin{equation}
D^{00}({\bf k})=-\frac{4\pi}{{\bf k}^2}, \quad D^{ij}({\bf k})=
-\frac{4\pi}{k^2}\left(\delta^{ij}-\frac{k^ik^j}{{\bf k}^2}\right),
\quad D^{0i}=D^{i0}=0,
\end{equation}
and ${\bf k=p-q}$; $\gamma_{\mu}$ and $u(p)$ are 
the Dirac matrices and spinors
\begin{equation}
\label{spinor}
u^\lambda({p})=\sqrt{\frac{\epsilon(p)+m}{2\epsilon(p)}}
\left(
\begin{array}{c}1\cr {\displaystyle\frac{\bm{\sigma}
      {\bf  p}}{\epsilon(p)+m}}
\end{array}\right)\chi^\lambda.
\end{equation}
Here  $\bm{\sigma}$   and $\chi^\lambda$
are the Pauli matrices and spinors; $\epsilon(p)=\sqrt{p^2+m^2}$.
The effective long-range vector vertex is
given by
\begin{equation}
\label{kappa}
\Gamma_{\mu}({\bf k})=\gamma_{\mu}+
\frac{i\kappa}{2m}\sigma_{\mu\nu}k^{\nu},
\end{equation}
where $\kappa$ is the Pauli interaction constant characterizing the
long-range anomalous chromomagnetic moment of quarks. Vector and
scalar confining potentials in the nonrelativistic limit reduce to
\begin{eqnarray}
\label{vlin}
V_V(r)&=&(1-\varepsilon)Ar+B,\nonumber\\ 
V_S(r)& =&\varepsilon Ar,
\end{eqnarray}
reproducing 
\begin{equation}
\label{nr}
V_{\rm conf}(r)=V_S(r)+V_V(r)=Ar+B,
\end{equation}
where $\varepsilon$ is the mixing coefficient. 

The expression for the quasipotential of the heavy quarkonia,
expanded in $v^2/c^2$ without and with retardation corrections to the
confining potential, can be found in Refs.~\cite{mass1} and 
\cite{efgbc,mass}, respectively. The 
structure of the spin-dependent interaction is in agreement with
the parameterization of Eichten and Feinberg \cite{ef}. The
quasipotential for the heavy quark interaction with a light antiquark
without employing the expansion in inverse powers of the light quark
mass is given in Ref.~\cite{hlm}.  
All the parameters of
our model like quark masses, parameters of the linear confining potential
$A$ and $B$, mixing coefficient $\varepsilon$ and anomalous
chromomagnetic quark moment $\kappa$ are fixed from the analysis of
heavy quarkonium masses \cite{mass1} and radiative
decays \cite{gf}. The quark masses
$m_b=4.88$ GeV, $m_c=1.55$ GeV, $m_s=0.50$ GeV, $m_{u,d}=0.33$ GeV and
the parameters of the linear potential $A=0.18$ GeV$^2$ and $B=-0.16$ GeV
have usual values of quark models.  The value of the mixing
coefficient of vector and scalar confining potentials $\varepsilon=-1$
has been determined from the consideration of the heavy quark expansion
for the semileptonic $B\to D$ decays
\cite{fg} and charmonium radiative decays \cite{gf}.
Finally, the universal Pauli interaction constant $\kappa=-1$ has been
fixed from the analysis of the fine splitting of heavy quarkonia ${
}^3P_J$- states \cite{mass1}. Note that the 
long-range  magnetic contribution to the potential in our model
is proportional to $(1+\kappa)$ and thus vanishes for the 
chosen value of $\kappa=-1$. It has been known for a long time 
that the correct reproduction of the
spin-dependent part of the quark-antiquark interaction requires 
either assuming  the scalar confinement or equivalently  introducing the
Pauli interaction with $\kappa=-1$ \cite{schn,mass1,mass} in the vector
confinement.

\section{Matrix elements of the electroweak current for
  $\bm{\lowercase{c\to s,d}}$ transitions} \label{mml}

In order to calculate the exclusive semileptonic decay rate of the
$B_c$ meson, it is necessary to determine the corresponding matrix
element of the  weak current between meson states.
In the quasipotential approach,  the matrix element of the weak current
$J^W_\mu=\bar q\gamma_\mu(1-\gamma_5)c$, associated with $c\to q$ ($q=s$
or $d$) transition, between a $B_c$ meson with mass $M_{B_c}$ and
momentum $p_{B_c}$ and a final meson $F$ ($F=B_s^{(*)}$ or
$B^{(*)}$) with mass $M_F$ and momentum $p_F$ takes the form \cite{f}
\begin{equation}\label{mxet} 
\langle F(p_F) \vert J^W_\mu \vert B_c(p_{B_c})\rangle
=\int \frac{d^3p\, d^3q}{(2\pi )^6} \bar \Psi_{F\,{\bf p}_F}({\bf
p})\Gamma _\mu ({\bf p},{\bf q})\Psi_{B_c\,{\bf p}_{B_c}}({\bf q}),
\end{equation}
where $\Gamma _\mu ({\bf p},{\bf
q})$ is the two-particle vertex function and  
$\Psi_{M\,{\bf p}_M}$ are the
meson ($M=B_c,F)$ wave functions projected onto the positive energy 
states of
quarks and boosted to the moving reference frame with momentum ${\bf p}_M$.
\begin{figure}
  \centering
  \includegraphics{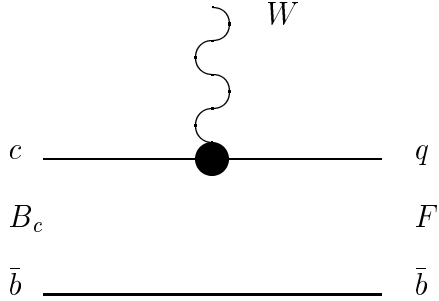}
\caption{Lowest order vertex function $\Gamma^{(1)}$
contributing to the current matrix element (\ref{mxet}). \label{d1}}
\end{figure}

\begin{figure}
  \centering
  \includegraphics{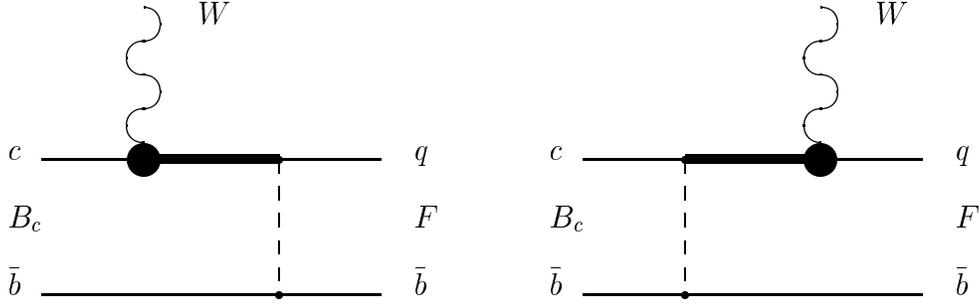}
\caption{ Vertex function $\Gamma^{(2)}$
taking the quark interaction into account. Dashed lines correspond  
to the effective potential ${\cal V}$ in 
(\ref{qpot}). Bold lines denote the negative-energy part of the quark
propagator. \label{d2}}
\end{figure}

 The contributions to $\Gamma$ come from Figs.~\ref{d1} and \ref{d2}. 
The contribution $\Gamma^{(2)}$ is the consequence
of the projection onto the positive-energy states. Note that the form of the
relativistic corrections resulting from the vertex function
$\Gamma^{(2)}$ is explicitly dependent on the Lorentz structure of the
quark-antiquark interaction. In the leading order of the $v^2/c^2$
expansion for 
$B_c$ and in the heavy quark limit $m_{b}\to \infty$ for $B_s,B$
only $\Gamma^{(1)}$ contributes, while $\Gamma^{(2)}$  
contributes already at the subleading order. 
The vertex functions look like
\begin{equation} \label{gamma1}
\Gamma_\mu^{(1)}({\bf
p},{\bf q})=\bar u_{q}(p_q)\gamma_\mu(1-\gamma^5)u_c(q_c)
(2\pi)^3\delta({\bf p}_b-{\bf q}_b),
\end{equation}
and
\begin{eqnarray}\label{gamma2} 
\Gamma_\mu^{(2)}({\bf p},{\bf q})&=&\bar u_{q}(p_q)\bar u_b(p_b)
\Bigl\{\gamma_{1\mu}(1-\gamma_1^5) 
\frac{\Lambda_c^{(-)}(
k)}{\epsilon_c(k)+\epsilon_c(p_q)}\gamma_1^0
{\cal V}({\bf p}_b-{\bf q}_b)\nonumber \\ 
& &+{\cal V}({\bf p}_b-{\bf q}_b)
\frac{\Lambda_{q}^{(-)}(k')}{ \epsilon_{q}(k')+
\epsilon_{q}(q_c)}\gamma_1^0 \gamma_{1\mu}(1-\gamma_1^5)\Bigr\}u_c(q_c)
u_b(q_b),\end{eqnarray}
where the superscripts ``(1)" and ``(2)" correspond to Figs.~\ref{d1} and
\ref{d2},  ${\bf k}={\bf p}_q-{\bf\Delta};\
{\bf k}'={\bf q}_c+{\bf\Delta};\ {\bf\Delta}={\bf
p}_F-{\bf p}_{B_c}$;
$$\Lambda^{(-)}(p)=\frac{\epsilon(p)-\bigl( m\gamma
^0+\gamma^0({\bm{ \gamma}{\bf p}})\bigr)}{ 2\epsilon (p)}.$$
Here \cite{f} 
\begin{eqnarray*} 
p_{q,b}&=&\epsilon_{q,b}(p)\frac{p_F}{M_F}
\pm\sum_{i=1}^3 n^{(i)}(p_F)p^i,\\
q_{c,b}&=&\epsilon_{c,b}(q)\frac{p_{B_c}}{M_{B_c}} \pm \sum_{i=1}^3 n^{(i)}
(p_{B_c})q^i,\end{eqnarray*}
and $n^{(i)}$ are three four-vectors given by
$$ n^{(i)\mu}(p)=\left\{ \frac{p^i}{M},\ \delta_{ij}+
\frac{p^ip^j}{M(E+M)}\right\}, \quad E=\sqrt{{\bf p}^2+M^2}.$$

It is important to note that the wave functions entering the weak current
matrix element (\ref{mxet}) are not in the rest frame in general. For example, 
in the $B_c$ meson rest frame (${\bf p}_{B_c}=0$), the final  meson
is moving with the recoil momentum ${\bf \Delta}$. The wave function
of the moving  meson $\Psi_{F\,{\bf\Delta}}$ is connected 
with the  wave function in the rest frame 
$\Psi_{F\,{\bf 0}}\equiv \Psi_F$ by the transformation \cite{f}
\begin{equation}
\label{wig}
\Psi_{F\,{\bf\Delta}}({\bf
p})=D_q^{1/2}(R_{L_{\bf\Delta}}^W)D_b^{1/2}(R_{L_{
\bf\Delta}}^W)\Psi_{F\,{\bf 0}}({\bf p}),
\end{equation}
where $R^W$ is the Wigner rotation, $L_{\bf\Delta}$ is the Lorentz boost
from the meson rest frame to a moving one, and   
the rotation matrix $D^{1/2}(R)$ in spinor representation is given by
\begin{equation}\label{d12}
{1 \ \ \,0\choose 0 \ \ \,1}D^{1/2}_{q,b}(R^W_{L_{\bf\Delta}})=
S^{-1}({\bf p}_{q,b})S({\bf\Delta})S({\bf p}),
\end{equation}
where
$$
S({\bf p})=\sqrt{\frac{\epsilon(p)+m}{2m}}\left(1+\frac{\bm{\alpha}{\bf p}}
{\epsilon(p)+m}\right)
$$
is the usual Lorentz transformation matrix of the four-spinor.

The general structure of the current matrix element (\ref{mxet}) is
rather complicated, because it is necessary to integrate both with
respect to $d^3p$ and $d^3q$. The $\delta$-function in the expression
(\ref{gamma1}) for the vertex function $\Gamma^{(1)}$ permits to perform
one of these integrations. As a result the contribution of
$\Gamma^{(1)}$ to the current matrix element has the usual structure of
an overlap integral of meson wave functions and
can be calculated exactly (without employing any expansion) in the
whole kinematical range, if the wave functions of the
initial and final mesons are known. The situation with the contribution
$\Gamma^{(2)}$ is different. Here, instead of a $\delta$-function, we have
a complicated structure, containing the potential of the $q\bar
q$-interaction in a meson. Thus in the general case we cannot accomplish one
of the integrations in the contribution of $\Gamma^{(2)}$ to the
matrix element (\ref{mxet}). Therefore, one should use some 
additional considerations in order to simplify calculations. The main
idea is to expand the vertex 
function $\Gamma^{(2)}$, given by (\ref{gamma2}), in such  a way that it
will be possible to use the quasipotential equation (\ref{quas}) in order
to perform one of the integrations in the current matrix element
(\ref{mxet}).  

The natural expansion parameters for $B_c$ decays to $B_s$, $B$ mesons
are the active $c$ and spectator $b$ quark masses. However, the
heavy $c$ quark undergoes the weak transition to the light $s$ or $d$
quark. The constituent  $s,d$ quark masses are of the same order of
magnitude as the relative momentum and binding energy, thus we cannot
apply the expansion in inverse powers of their masses. The heavy quark
expansion in $1/m_{c,b}$ significantly simplifies the structure of
the $\Gamma^{(2)}$ contribution to the decay matrix element, but the
momentum ${\bf p}$ dependence of the light quark energies
$\epsilon_q(p)$ still prevents to perform one of the
integrations. It is important to note that the kinematically allowed
range for $B_c$ decays to $B_s$ and $B$ meson is not large 
($|{\bf\Delta}_{\rm max}|=(M_{B_c}^2-M_F^2)/(2M_{B_c})$ $\sim 0.8$~GeV for
decays to $B_s$ and $\sim 0.9$~GeV for decays to $B$ mesons). This
means that the recoil momentum ${\bf \Delta}$ of a final meson is
of the same order as the relative momentum ${\bf p}$ of
quarks inside a heavy-light meson ($\sim 0.5$~GeV) in the whole
kinematical range. Taking also into account that the
final $B_s$ and $B$ mesons are weakly bound \cite{hlm}, 
\footnote{The sum of constituent quark  masses $m_b+m_q$ is very close
  to the ground state meson mass $M$.} we can replace the light quark
energies by the center of mass energies on mass shell
$\epsilon_q(p)\to E_q=(M_F^2-m_b^2+m_q^2)/(2M_F)$. We used such a
substitution in our analysis of heavy-light meson mass spectra
\cite{hlm} which allowed us to treat the light quark relativistically
without an unjustified expansion in inverse powers of its mass. Making
these replacements and expansions we see that it is possible to 
integrate the current matrix element (\ref{mxet}) either with
respect to $d^3p$ or $d^3q$ using the quasipotential equation
(\ref{quas}). Performing integrations and
taking the sum of the contributions $\Gamma^{(1)}$ and
$\Gamma^{(2)}$ we get the expression for the current matrix element,
which contains ordinary overlap integrals of meson wave functions
and is valid in the whole kinematical range.
Hence the matrix element can be easily calculated using numerical wave
functions found in our analysis of the meson mass spectra
\cite{efgbc,mass}.             

\section{$\bm{B_{\lowercase{c}}}$ decay form factors}\label{dff}

The matrix elements of the weak current $J^W$ for $B_c$ decays
to pseudoscalar  mesons
($P=B_s, B$) can be parametrized by two invariant form factors:
\begin{equation}
  \label{eq:pff1}
  \langle P(p_F)|\bar q \gamma^\mu c|B_c(p_{B_c})\rangle
  =f_+(q^2)\left[p_{B_c}^\mu+ p_F^\mu-
\frac{M_{B_c}^2-M_P^2}{q^2}\ q^\mu\right]+
  f_0(q^2)\frac{M_{B_c}^2-M_P^2}{q^2}\ q^\mu,
\end{equation}
where $q=p_{B_c}-p_F$; $M_{B_c}$ is the $B_c$ meson mass 
and $M_P$ is the pseudoscalar meson mass. 

The corresponding matrix elements for $B_c$  decays to vector mesons
($V=B_s^*, B^*$) are parametrized by four form factors
\begin{eqnarray}
  \label{eq:vff1}
  \langle V(p_F)|\bar q \gamma^\mu c|B(p_{B_c})\rangle&=
  &\frac{2iV(q^2)}{M_{B_c}+M_V} \epsilon^{\mu\nu\rho\sigma}\epsilon^*_\nu
  p_{B_c\rho} p_{F\sigma},\\ \cr
\label{eq:vff2}
\langle V(p_F)|\bar q \gamma^\mu\gamma_5 c|B(p_{B_c})\rangle&=&2M_V
A_0(q^2)\frac{\epsilon^*\cdot q}{q^2}\ q^\mu
 +(M_{B_c}+M_V)A_1(q^2)\left(\epsilon^{*\mu}-\frac{\epsilon^*\cdot
    q}{q^2}\ q^\mu\right)\cr\cr
&&-A_2(q^2)\frac{\epsilon^*\cdot q}{M_{B_c}+M_V}\left[p_{B_c}^\mu+
  p_F^\mu-\frac{M_{B_c}^2-M_V^2}{q^2}\ q^\mu\right], 
\end{eqnarray}
where 
$M_V$ and $\epsilon_\mu$ are the mass and polarization vector of
the final vector meson. The following relations hold for the form
factors at the maximum recoil point of the final meson ($q^2=0$)
\[f_+(0)=f_0(0),\]
\[A_0(0)=\frac{M_{B_c}+M_V}{2M_V}A_1(0)
-\frac{M_{B_c}-M_V}{2M_V}A_2(0).\]
In the limit of vanishing lepton mass, the form factors $f_0$ and $A_0$ do
not contribute to the semileptonic decay rates. However, they
contribute to nonleptonic decay rates in the factorization
approximation.  

It is convenient to consider $B_c$ semileptonic and nonleptonic decays
in the $B_c$ meson rest frame. Then it is important to take into account 
the boost of the final meson wave function from the rest reference
frame to the moving one with the recoil momentum ${\bf \Delta}$, given
by Eq.~(\ref{wig}). Now we can apply the method for
calculating decay matrix elements described in the previous section.   
As it is argued above, the leading contributions arising from the vertex
function $\Gamma^{(1)}$ can be exactly expressed through the overlap
integrals of the meson wave functions in the whole kinematical range. 
For the subleading contribution $\Gamma^{(2)}$, the expansion in powers
of the ratio of the relative quark momentum ${\bf p}$ to heavy quark
masses $m_{b,c}$ should be performed. Taking into account that the
recoil momentum of the final meson ${\bf \Delta}$ is not large we
replace the final light quark energies $\epsilon_q(p)$ by the center
of mass energies on mass shell $E_q$. Such replacement is well
justified near the point of zero recoil of the final $B_s$, $B$ meson.
The weak dependence  of this subleading contribution on the recoil
momentum and its numerical smallness due to its proportionality to
the small meson binding energy  permits its extrapolation to the whole
kinematical range. 
As a result, we get the following expressions for the $B_c$ decay
form factors:

(a) $B_c\to P$ transitions ($P=B_s,B$) 
\begin{equation}
  \label{eq:f+}
  f_+(q^2)=f_+^{(1)}(q^2)+\varepsilon f_+^{S(2)}(q^2)
+(1-\varepsilon) f_+^{V(2)}(q^2),
\end{equation}
\begin{equation}
  \label{eq:f0}
  f_0(q^2)=f_0^{(1)}(q^2)+\varepsilon f_0^{S(2)}(q^2)
+(1-\varepsilon) f_0^{V(2)}(q^2),
\end{equation}

(b) $B_c\to V$ transition ($V=B_s^*,B^*$)
\begin{equation}
  \label{eq:V}
  V(q^2)=V^{(1)}(q^2)+\varepsilon V^{S(2)}(q^2)
+(1-\varepsilon) V^{V(2)}(q^2),
\end{equation}
\begin{equation}
  \label{eq:A1}
  A_1(q^2)=A_1^{(1)}(q^2)+\varepsilon A_1^{S(2)}(q^2)
+(1-\varepsilon) A_1^{V(2)}(q^2),
\end{equation}
\begin{equation}
  \label{eq:A2}
  A_2(q^2)=A_2^{(1)}(q^2)+\varepsilon A_2^{S(2)}(q^2)
+(1-\varepsilon) A_2^{V(2)}(q^2),
\end{equation}
\begin{equation}
  \label{eq:A0}
  A_0(q^2)=A_0^{(1)}(q^2)+\varepsilon A_0^{S(2)}(q^2)
+(1-\varepsilon) A_0^{V(2)}(q^2),
\end{equation}
where $f_{+,0}^{(1)}$, $f_{+,0}^{S,V(2)}$, $A_{0,1,2}^{(1)}$,
$A_{0,1,2}^{S,V(2)}$, $V^{(1)}$ and $V^{S,V(2)}$ are given in Appendix.
The superscripts ``(1)" and ``(2)" correspond to Figs.~\ref{d1} and
\ref{d2}, $S$ and
$V$ to the scalar and vector potentials of $q\bar q$-interaction.
The mixing parameter of scalar and vector confining potentials
$\varepsilon$ is fixed to be $-1$  in our model.

It is easy to check that in the heavy quark limit the decay matrix
elements (\ref{eq:pff1})--(\ref{eq:vff2}) with form factors
(\ref{eq:f+})--(\ref{eq:A0}) satisfy the heavy quark spin symmetry
relations \cite{jlms} obtained near the zero recoil point (${\bf
  \Delta}\to 0$).

\begin{table}
\caption{Form factors of weak $B_c$ decays ($c\to s(d)$ transitions). }
\label{ff}
\begin{ruledtabular}
\begin{tabular}{ccccccc}
Transition   & $f_+(q^2)$ & $f_0(q^2)$ & $V(q^2)$ & $A_1(q^2)$ 
& $A_2(q^2)$ &  $A_0(q^2)$\\
\hline
$B_c\to B_s(B_s^*)$ & \\
$q^2=q^2_{\rm max}$ & 0.99 & 0.86 & 6.25 & 0.76 & 2.62 & 0.91\\  
$q^2=0$ & 0.50 & 0.50 & 3.44 & 0.49 & 2.19 & 0.35\\
$B_c\to B(B^*)$ &\\
$q^2=q^2_{\rm max}$ & 0.96 & 0.80 & 8.91 & 0.72 & 2.83 & 1.06   \\
$q^2=0$ & 0.39 & 0.39 & 3.94 & 0.42 & 2.89 & 0.20 \\
\end{tabular}
\end{ruledtabular}
\end{table}

\begin{figure}
  \centering
 \includegraphics{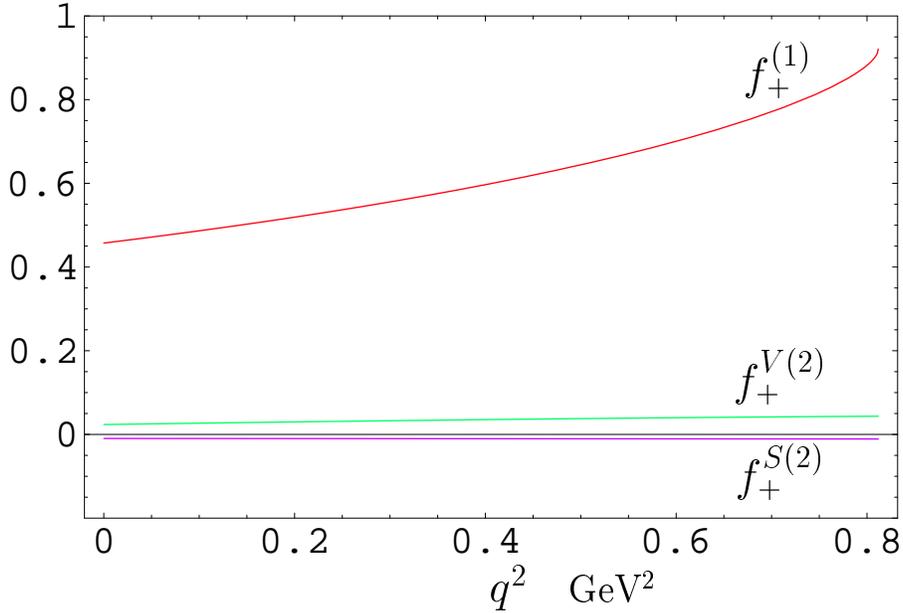}
\caption{Leading $f_+^{(1)}$ and subleading $f_+^{S(2)}$, $f_+^{V(2)}$
  contributions to the form factor $f_+$ for the $B_c\to B_s$ transition.}
  \label{fpbcbs}  
\end{figure}

For numerical calculations we use the quasipotential wave functions of the
$B_c$ meson, $B_s$ and $B$ mesons obtained in the mass spectrum
calculations \cite{mass,hlm}.  Our model predicts the $B_c$ meson mass  
$M_{B_c}=6.270$~GeV \cite{efgbc}, while for $B_s^{(*)}$ and $B^{(*)}$
meson masses we use experimental data \cite{pdg}.
The calculated values of form factors at zero 
($q^2=q^2_{\rm max}$) and maximum ($q^2=0$) recoil of the final meson
are listed in Table~\ref{ff}. In Fig.~\ref{fpbcbs} we plot leading
$f_+^{(1)}$ and subleading $f_+^{S(2)}$, $f_+^{V(2)}$ contributions to the
form factor $f_+$ for the $B_c\to B_s$ transition, as an example. We see
that the leading contribution $f_+^{(1)}$ is dominant in the whole
kinematical range, as it was expected. The subleading contributions
$f_+^{S(2)}$, $f_+^{V(2)}$ are small and weakly depend on $q^2$. The
behavior of corresponding contributions to other form factors is
similar. This supports our conjecture that the formulae 
(\ref{eq:fpl})--(\ref{eq:a0v}) can be applied for the calculation of
the form factors of $B_c\to B_s(B)^{(*)}$ transitions in the whole
kinematical range.

\begin{figure}
  \centering
  \includegraphics{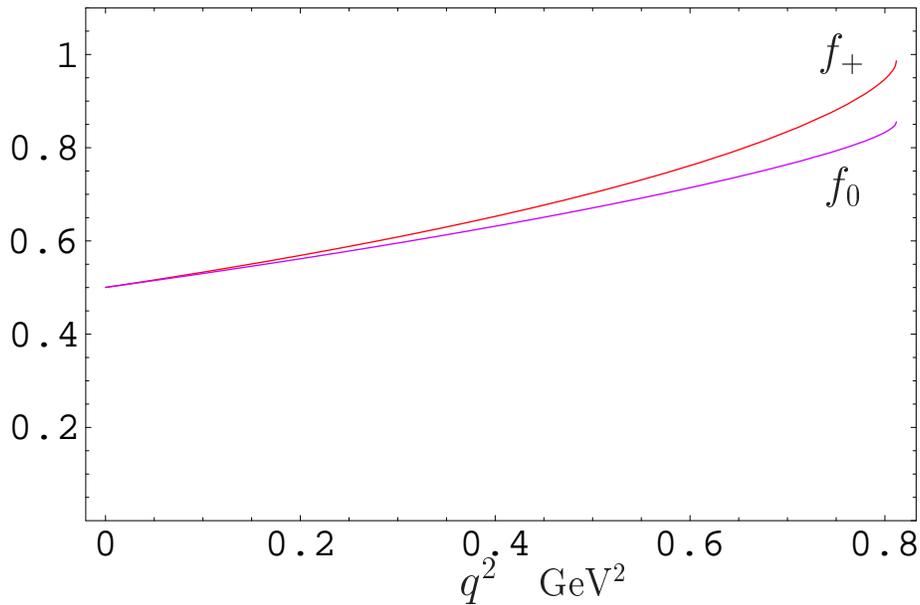}
\caption{Form factors of the $B_c\to B_s e\nu$ decay.} \label{bctobsff}
\end{figure}

\begin{figure}
  \centering
  \includegraphics{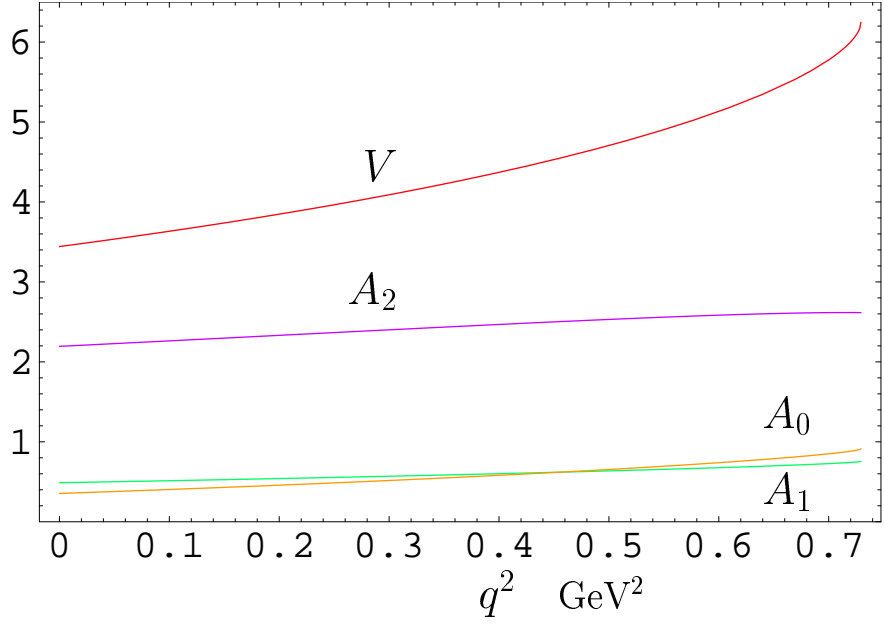}
\caption{Form factors of the $B_c\to B_s^* e\nu$ decay.}\label{bctobsvff} 
\end{figure}

\begin{figure}
  \centering
  \includegraphics{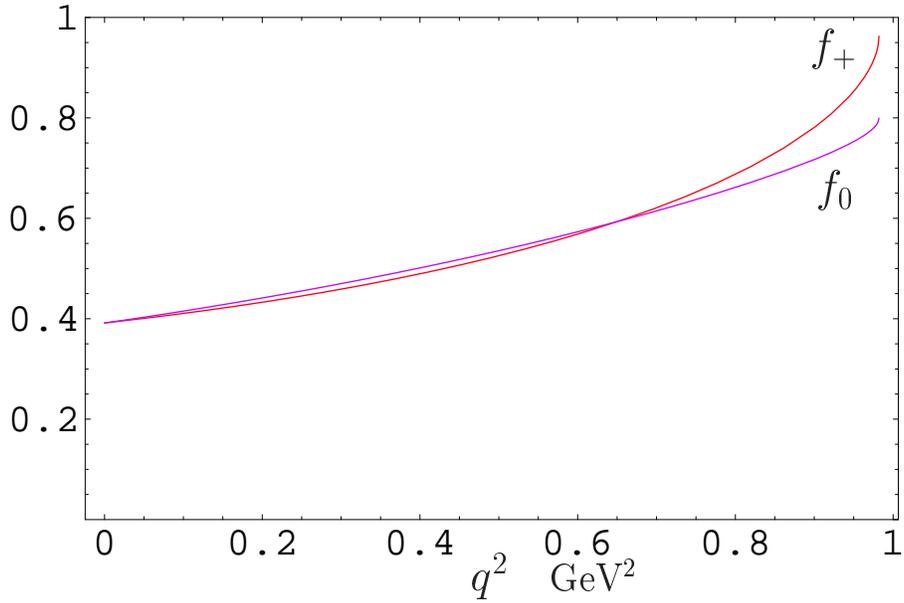}
\caption{Form factors of the $B_c\to B e\nu$ decay.}\label{bctobff}
\end{figure}

\begin{figure}
  \centering
  \includegraphics{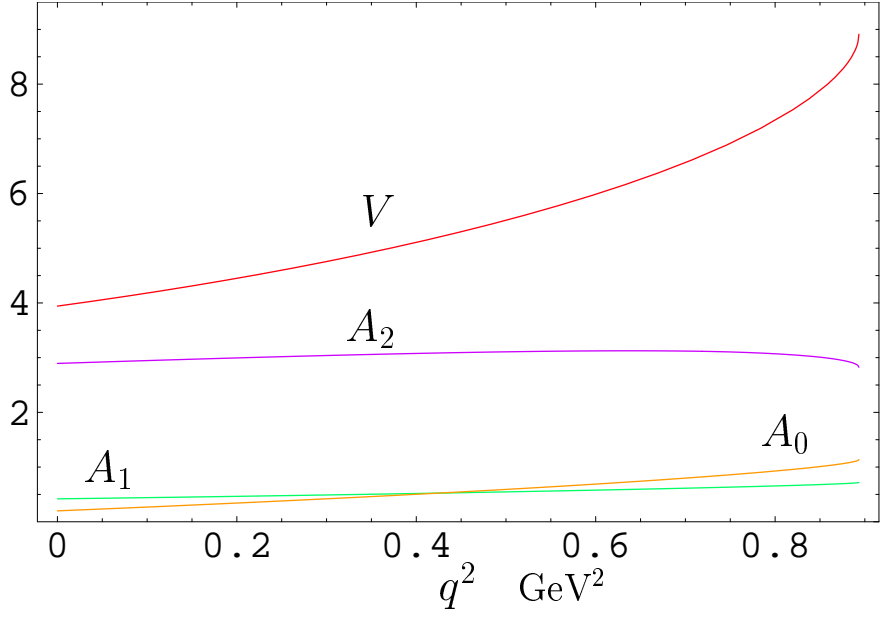}
\caption{Form factors of the $B_c\to B^* e\nu$ decay.}\label{bctobvff}
\end{figure}

In Figs.~\ref{bctobsff}-\ref{bctobvff} we plot the calculated $q^2$
dependence of the weak form factors of Cabibbo-Kobayashi-Maskawa (CKM)
favored ($B_c\to B_s$, $B_c\to B_s^*$), as well as CKM suppressed
($B_c\to B$, $B_c\to B^*$) transitions in the whole kinematical range. 
  
In the following sections we use the obtained form factors for the
calculation of the semileptonic and nonleptonic $B_c$ decay rates.

\section{Semileptonic decays}\label{ssd}

The differential semileptonic decay rates can be expressed in terms of
the form factors as follows.

(a) $B_c\to Pe\nu$ decays  ($P=B_s,B$)
\begin{equation}
  \label{eq:dgp}
  \frac{{\rm d}\Gamma}{{\rm d}q^2}(B_c\to Pe\nu)=\frac{G_F^2 
  \Delta^3 |V_{cq}|^2}{24\pi^3} |f_+(q^2)|^2.
\end{equation}

(b) $B_c\to Ve\nu$ decays ($V=B_s^*,B^*$)
\begin{equation}
  \label{eq:dgv}
\frac{{\rm d}\Gamma}{{\rm d}q^2}(B_c\to Ve\nu)=\frac{G_F^2
\Delta|V_{cq}|^2}{96\pi^3}\frac{q^2}{M_{B_c}^2}
\left(|H_+(q^2)|^2+|H_-(q^2)|^2   
+|H_0(q^2)|^2\right),
\end{equation}
where $G_F$ is the Fermi constant, $V_{cq}$ is the
CKM matrix element ($q=s,d$),
\[\Delta\equiv|{\bf\Delta}|=\sqrt{\frac{(M_{B_c}^2+M_{P,V}^2-q^2)^2}
{4M_{B_c}^2}-M_{P,V}^2}.
\]
The helicity amplitudes are given by
\begin{equation}
  \label{eq:helamp}
  H_\pm(q^2)=\frac{2M_{B_c}\Delta}{M_{B_c}+M_V}\left[V(q^2)\mp
\frac{(M_{B_c}+M_V)^2}{2M_{B_c}\Delta}A_1(q^2)\right],
\end{equation}
\begin{equation}
  \label{eq:h0a}
  H_0(q^2)=\frac1{2M_V\sqrt{q^2}}\left[(M_{B_c}+M_V)
(M_{B_c}^2-M_V^2-q^2)A_1(q^2)-\frac{4M_{B_c}^2\Delta^2}{M_{B_c}
+M_V}A_2(q^2)\right].
\end{equation}
The decay rates to the longitudinally and transversely polarized
vector mesons are defined by
\begin{equation}
  \label{eq:dgl}
\frac{{\rm d}\Gamma_L}{{\rm d}q^2}=\frac{G_F^2
\Delta|V_{cq}|^2}{96\pi^3}\frac{q^2}{M_{B_c}^2}
|H_0(q^2)|^2,  
\end{equation}
\begin{equation}
  \label{eq:dgt}
\frac{{\rm d}\Gamma_T}{{\rm d}q^2}=
\frac{{\rm d}\Gamma_+}{{\rm d}q^2}+\frac{{\rm d}\Gamma_-}{{\rm d}q^2}
=\frac{G_F^2\Delta|V_{cq}|^2}{96\pi^3}\frac{q^2}{M_{B_c}^2}
\left(|H_+(q^2)|^2+|H_-(q^2)|^2\right). 
\end{equation}

\begin{figure}
  \centering
\vskip 1.5cm
  \includegraphics[scale=1]{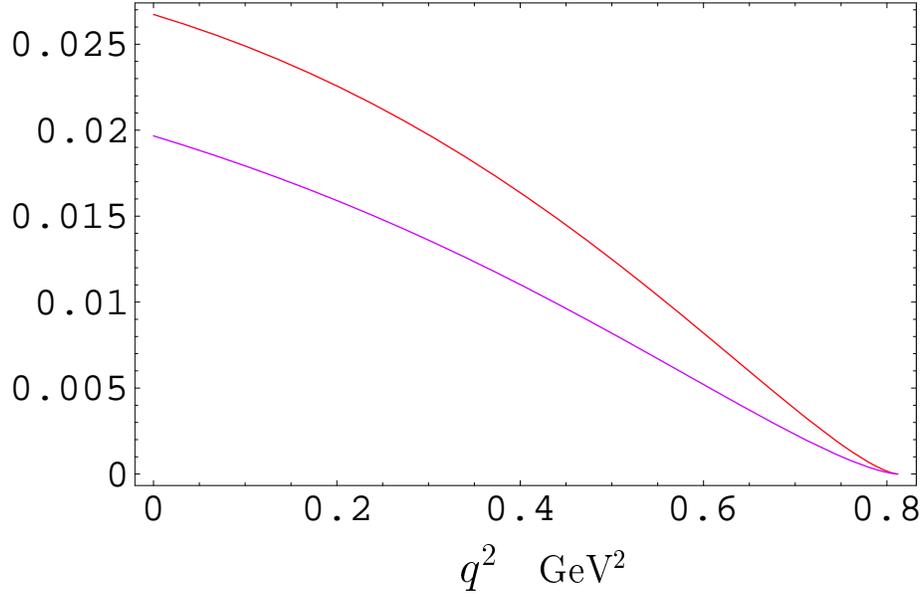}


\caption{Differential decay rates
  $(1/{|V_{cs}|^2}){{\rm d}\Gamma}/{{\rm d}q^2}$ of $B_c\to B_s e\nu$
  decay (in $10^{-12}$~GeV$^{-1}$). The lower curve is evaluated without
  account of $1/m_{b,c}$ corrections.} \label{dgbcbs}
\end{figure}

\begin{figure}
  \centering\vskip 0.1cm
  \includegraphics[scale=1]{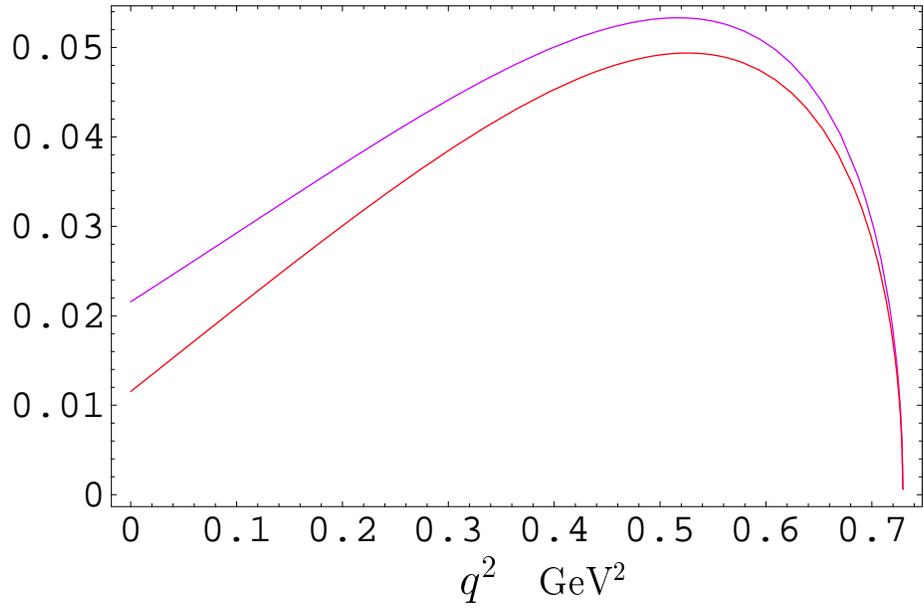}


\caption{Differential decay rates
  $(1/{|V_{cs}|^2}){{\rm d}\Gamma}/{{\rm d}q^2}$ of $B_c\to B_s^* e\nu$
  decay (in $10^{-12}$~GeV$^{-1}$). The upper curve is
  evaluated without account of $1/m_{b,c}$ corrections.} \label{dgbcbsv}
\end{figure}

\begin{figure}
  \centering
\vskip 1.5cm
  \includegraphics[scale=1]{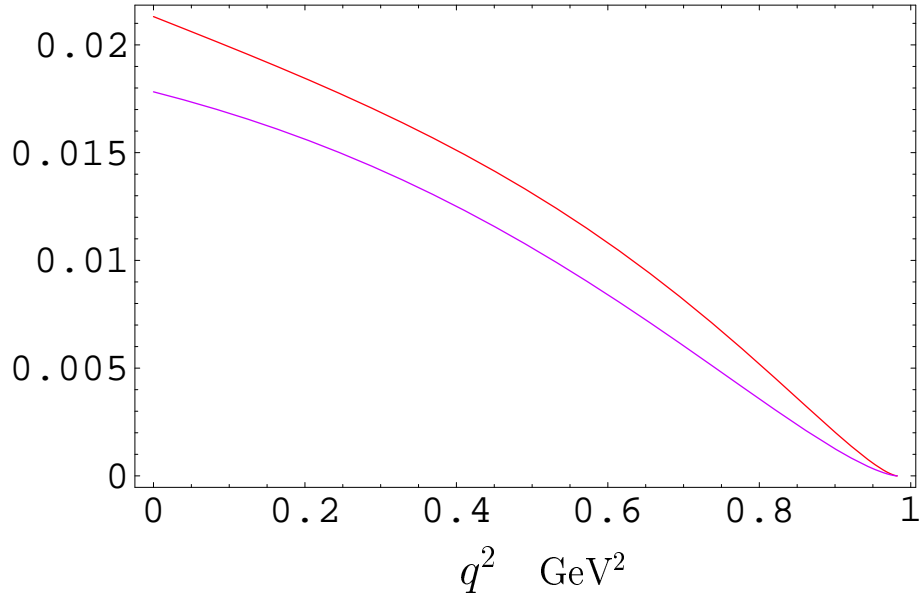}


\caption{Differential decay rate
  $(1/{|V_{cd}|^2}){{\rm d}\Gamma}/{{\rm d}q^2}$ of $B_c\to B e\nu$
  decay (in $10^{-12}$ GeV$^{-1}$). The lower curve is evaluated without
  account of $1/m_{b,c}$ corrections.}\label{dgbcb}
\end{figure}

\begin{figure}
  \centering\vskip 0.1cm
  \includegraphics[scale=1]{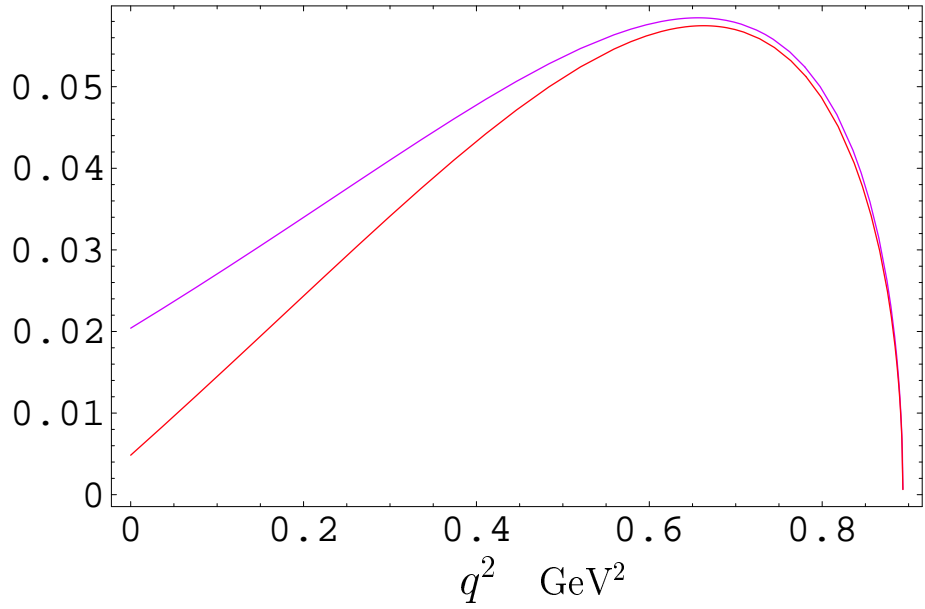}


\caption{Differential decay rates
  $(1/{|V_{cd}|^2}){{\rm d}\Gamma}/{{\rm d}q^2}$ of $B_c\to B^* e\nu$
  decay (in $10^{-12}$ GeV$^{-1}$). The upper  curve is
  evaluated without account of $1/m_{b,c}$ corrections.}\label{dgbcbv}

\end{figure}

In Figs.~\ref{dgbcbs}-\ref{dgbcbv} we plot the differential
semileptonic decay rates ${\rm d}\Gamma/{{\rm d}q^2}$ for semileptonic
decays $B_c\to B_s^{(*)}e\nu$ and
$B_c\to B^{(*)}e\nu$  calculated in our model using
Eqs.~(\ref{eq:dgp}), (\ref{eq:dgv}) both with and without
account of $1/m_{b,c}$ corrections to the decay form factors
(\ref{eq:fpl})--(\ref{eq:a0v}).~\footnote{Relativistic wave functions
  were used for both calculations.}   From these plots we see that
relativistic effects related to heavy quarks increase the rates of
semileptonic $B_c$ decays 
to the pseudoscalar $B_s$ and $B$ mesons and decrease the rates of
semileptonic decays to vector $B_s^*$ and $B^*$ mesons.

\begin{table}
\caption{Semileptonic decay rates $\Gamma$ of $B_c$ to  $B_s^{(*)}$
   and $B^{(*)}$ mesons (in $10^{-15}$ GeV). } 
\label{sdr}
\begin{ruledtabular}
\begin{tabular}{ccccccccccc}
Decay& our& \cite{iks}& \cite{klo} & \cite{emv} & \cite{cc} &
\cite{cdf} & \cite{aknt} & \cite{nw} &\cite{lyl}&\cite{lc}\\
\hline
$B_c\to B_s e\nu$ & 12 & 29 & 59 & 14.3 & 26.6 & 11.1(12.9) & 15 &
12.3 & 11.75 & 26.8\\
$B_c\to B_s^* e\nu$ & 25 & 37 & 65 & 50.4 & 44.0 & 33.5(37.0) &
34 & 19.0 & 32.56 & 34.6\\
$B_c\to B e\nu$ & 0.6 & 2.1 & 4.9 & 1.14 & 2.30 & 0.9(1.0) &
& & 0.59 & 1.90\\
$B_c\to B^* e\nu$ & 1.7 & 2.3 & 8.5 & 3.53 & 3.32 & 2.8(3.2) & &
& 2.44 & 2.34
\end{tabular}
\end{ruledtabular}
\end{table}

We calculate the total rates of the semileptonic $B_c$ decays  
to the $B_s^{(*)}$ and $B^{(*)}$ mesons integrating the corresponding 
differential decay rates over $q^2$. For calculations we use the
following values of the CKM matrix elements: $|V_{cs}|=0.974$,
$|V_{cd}|=0.223$.  The results  are given in
Table~\ref{sdr} in comparison with predictions of other
approaches based on 
quark models \cite{iks,emv,cc,aknt,nw,lc}, QCD sum
rules \cite{klo} and on the application of heavy quark
symmetry relations \cite{cdf,lyl} to the quark model. Our predictions
for the CKM favored semileptonic $B_c$ decays to $B_s^{(*)}$
are  smaller than those of QCD sum rules \cite{klo} and
quark models \cite{iks,emv,cc,lc}, but agree with quark model results
\cite{cdf,aknt,nw,lyl}. For the CKM suppressed
semileptonic decays of $B_c$ to $B^{(*)}$ mesons our results are in
agreement with the ones based on the application of heavy quark symmetry
relations \cite{cdf,lyl} to the quark model. 
     
In Table~\ref{sr} we present for completeness our predictions for the rates
of the semileptonic $B_c$ decays  to vector ($B_s^*$ and $B^*$) mesons
with longitudinal ($L$) or transverse ($T$) polarization and to the
states with helicities $\lambda=\pm 1$, as well as their ratios.

\begin{table}
\caption{Semileptonic decay rates $\Gamma_{L,T,+,-}$ (in $10^{-15}$
  GeV) and their ratios for $B_c$ decays to vector $B_s^*$ and $B^*$
  mesons.} 
\label{sr}
\begin{ruledtabular}
\begin{tabular}{ccccccc}
Decay& $\Gamma_L$ & $\Gamma_T$ & $\Gamma_L/\Gamma_T$ & $\Gamma_+$ &
$\Gamma_-$ &  $\Gamma_+/\Gamma_-$\\
\hline
$B_c\to B_s^* e\nu$ & 10.5 & 14.5 & 0.74 & 3.1 & 11.4 & 0.27\\
$B_c\to B^* e\nu$ & 0.57 & 1.13 & 0.50 & 0.13 & 1.0 & 0.13
\end{tabular}
\end{ruledtabular}
\end{table}

\section{Nonleptonic decays}\label{nl}

In the standard model nonleptonic $B_c$ decays are described by the
effective Hamiltonian, obtained by integrating out the heavy $W$-boson
and top quark. For the case of $c\to s,d$ transitions, one gets
\begin{equation}
\label{heff}
H_{\rm eff}=\frac{G_F}{\sqrt{2}}V_{cs}\left[c_1(\mu)O_1^{cs}+
c_2(\mu)O_2^{cs}\right] 
+\frac{G_F}{\sqrt{2}}V_{cd}\left[c_1(\mu)O_1^{cd}+
c_2(\mu)O_2^{cd}\right] +\dots.
\end{equation}
The Wilson coefficients $c_{1,2}(\mu)$ are evaluated
perturbatively at the $W$ scale and then are evolved down to the
renormalization scale $\mu\approx m_c$ by the renormalization-group
equations. The ellipsis denote the penguin operators, the Wilson
coefficients of  which are numerically much smaller than $c_{1,2}$.
The local four-quark operators $O_1$ and $O_2$ are given by
\begin{eqnarray}
\label{o12}
O_1^{cq}&=& ({\tilde d}u)_{V-A}(\bar c
q)_{V-A}, \cr O_2^{cq}&=& (\bar
cu)_{V-A}({\tilde d}q)_{V-A}, \qquad q=(s,d),
\end{eqnarray}
where the rotated antiquark field is
\begin{equation} \label{ds}
\tilde d=V_{ud}\bar d+V_{us}\bar s
\end{equation}
and for
the hadronic current the following notation is used
$$(\bar qq')_{V-A}=\bar q\gamma_\mu(1-\gamma_5)q' \equiv J^W_\mu.$$

The factorization approach, which is extensively used for the calculation
of two-body nonleptonic decays, such as $B_c\to FM$, assumes that the
nonleptonic decay amplitude reduces to the product of a form factor
and a decay constant \cite{bsw}. This assumption in general cannot be
exact.  However, it is expected that factorization can hold 
for the energetic decays, where the final $F$ meson is heavy and the $M$
meson is light \cite{dg}. A justification of this assumption is
usually based on the issue of color 
transparency \cite{jb}. In these decays the final hadrons
are produced in the form of point-like color-singlet objects with a
large relative momentum. And thus the hadronization of the decay
products occurs  after they are too far away for strongly interacting
with each other. That provides the possibility to avoid the final
state interaction. A more general treatment of factorization is given in
Refs.~\cite{bbns,bs}.  

\begin{figure}
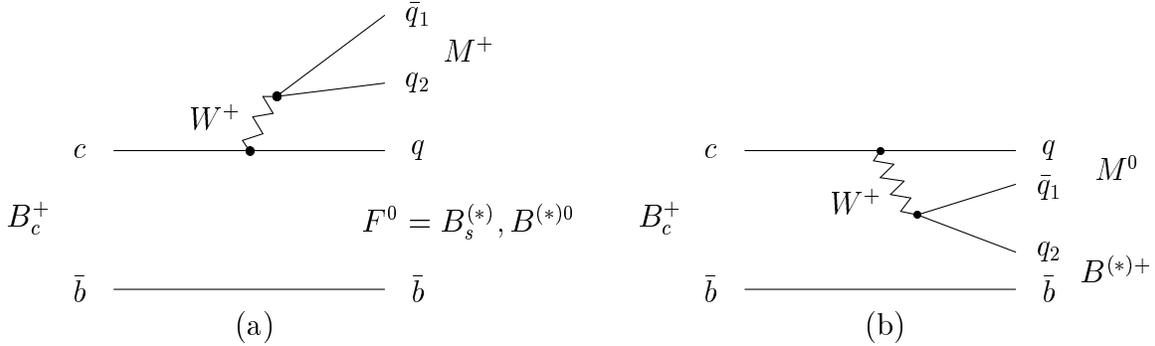

  \centering 
  \includegraphics{fig.12a}\qquad\includegraphics{fig.12b}
\caption{Quark diagrams for the nonleptonic $B_c$ decays: (a) $B_c^+\to
  F^0M^+$ decay; (b) $B_c^+\to B^+M^0$ decay. }
\label{d3}
\end{figure}

In this paper we consider the following two types of nonleptonic decays: 
(a) $B_c^+\to B_s^{(*)}(B^{(*)0})M^+$ and (b) $B_c^+\to B^{(*)+}M^0$,
where the final light $M^+$ and $M^0$ mesons are
$\pi$, $\rho$ or $K^{(*)}$. The corresponding diagrams are shown in
Fig.~\ref{d3}, where $q,q_1=d$,  $s$ and $q_2=u$. Then in the
factorization approximation the decay
amplitudes can be expressed through the product of one-particle matrix
elements 
\begin{eqnarray}
\label{factor} 
\langle F^0M^+|H_{\rm eff}|B_c^+\rangle&=& \frac{G_F}{\sqrt{2}}
V_{cq}V_{q_1q_2} a_1\langle F|(\bar
cq)_{V-A}|B_c\rangle\langle M|(\bar q_1q_2)_{V-A}|0\rangle ,\cr
\langle B^{(*)+}M^0|H_{\rm eff}|B_c^+\rangle&=& \frac{G_F}{\sqrt{2}}
V_{cq}V_{q_1q_2} a_2\langle B^{(*)}|(\bar
cq_2)_{V-A}|B_c\rangle\langle  M |(\bar q_1q)_{V-A}|0\rangle ,
\end{eqnarray}
where
\begin{equation}
\label{amu} a_1=c_1(\mu)+\frac{1}{N_c}c_2(\mu),
\qquad a_2=c_2(\mu)+\frac1{N_c}c_1(\mu)
\end{equation}
and $N_c$ is the number of colors. 

The matrix element of the current $J^W_\mu$ between vacuum and final
pseudoscalar ($P$) or vector ($V$) meson is parametrized by the decay
constants $f_{P,V}$
\begin{equation}
\langle P|\bar q_1 \gamma^\mu\gamma_5 q_2|0\rangle=if_Pp^\mu_P, \qquad
\langle V|\bar q_1\gamma_\mu q_2|0\rangle=\epsilon_\mu M_Vf_V.
\end{equation}
We use the following values of the decay constants: $f_\pi=0.131$~GeV,
$f_\rho=0.208$~GeV, $f_K=0.160$~GeV and $f_{K^*}=0.214$~GeV. The CKM
matrix elements are $|V_{ud}|=0.975$,  $|V_{us}|=0.222$.

\begin{table}[htb]
\caption{Nonleptonic decay rates $\Gamma$ (in $10^{-15}$ GeV). }
\label{ssdr}
\begin{ruledtabular}
\begin{tabular}{cccccccc}
Decay& our&  \cite{klo} & \cite{emv} & \cite{cc} & \cite{aknt} &
\cite{cdf}& \cite{lc}\\
\hline
$B_c^+\to B_s\pi^+$ & 25$a_1^2$ & 167$a_1^2$ & 15.8$a_1^2$ &
58.4$a_1^2$ & 34.8$a_1^2$ & 30.6$a_1^2$ & 65.1$a_1^2$ \\
$B_c^+\to B_s\rho^+$ & 14$a_1^2$ & 72.5$a_1^2$ & 39.2$a_1^2$ &
44.8$a_1^2$ & 23.6$a_1^2$ & 13.6$a_1^2$& 42.7$a_1^2$\\
$B_c^+\to B_s^*\pi^+$ & 16$a_1^2$ & 66.3$a_1^2$ & 12.5$a_1^2$ &
51.6$a_1^2$ & 19.8$a_1^2$ & 35.6$a_1^2$& 25.3$a_1^2$\\
$B_c^+\to B_s^*\rho^+$ & 110$a_1^2$ & 204$a_1^2$ & 171$a_1^2$ &
150$a_1^2$ & 123$a_1^2$ & 110.1$a_1^2$ & 139.6$a_1^2$\\
$B_c^+\to B_s K^+$ & 2.1$a_1^2$ & 10.7$a_1^2$ & 1.70$a_1^2$ &
4.20$a_1^2$ &  & 2.15$a_1^2$ & 4.69$a_1^2$\\
$B_c^+\to B_s K^{*+}$ & 0.03$a_1^2$ &  & 1.06$a_1^2$ &
 &  & 0.043$a_1^2$ & 0.296$a_1^2$\\
$B_c^+\to B_s^* K^+$ & 1.1$a_1^2$ & 3.8$a_1^2$ & 1.34$a_1^2$ &
2.96$a_1^2$ &  & 1.6$a_1^2$ & 1.34$a_1^2$\\
$B_c^+\to B^0\pi^+$ & 1.0$a_1^2$ & 10.6$a_1^2$ & 1.03$a_1^2$ &
3.30$a_1^2$ & 1.50$a_1^2$ & 1.97$a_1^2$& 3.64$a_1^2$\\
$B_c^+\to B^0\rho^+$ & 1.3$a_1^2$ & 9.7$a_1^2$ & 2.81$a_1^2$ &
5.97$a_1^2$ & 1.93$a_1^2$ & 1.54$a_1^2$ & 4.03$a_1^2$\\
$B_c^+\to B^{*0}\pi^+$ & 0.26$a_1^2$ & 9.5$a_1^2$ & 0.77$a_1^2$ &
2.90$a_1^2$ & 0.78$a_1^2$ & 2.4$a_1^2$& 1.22$a_1^2$\\
$B_c^+\to B^{*0}\rho^+$ & 6.8$a_1^2$ & 26.1$a_1^2$ & 9.01$a_1^2$ &
11.9$a_1^2$ & 6.78$a_1^2$ & 8.6$a_1^2$ & 8.16$a_1^2$\\
$B_c^+\to B^0 K^+$ & 0.09$a_1^2$ & 0.70$a_1^2$ & 0.105$a_1^2$ &
0.255$a_1^2$ &  & 0.14$a_1^2$ & 0.272$a_1^2$\\
$B_c^+\to B^0 K^{*+}$ & 0.04$a_1^2$ & 0.15$a_1^2$ & 0.125$a_1^2$ &
0.180$a_1^2$ &  & 0.032$a_1^2$ & 0.0965$a_1^2$\\
$B_c^+\to B^{*0} K^+$ & 0.04$a_1^2$ & 0.56$a_1^2$ & 0.064$a_1^2$ &
0.195$a_1^2$ &  & 0.12$a_1^2$ & 0.0742$a_1^2$\\
$B_c^+\to B^{*0} K^{*+}$ & 0.33$a_1^2$ & 0.59$a_1^2$ & 0.665$a_1^2$ &
0.374$a_1^2$ &  & 0.34$a_1^2$ & 0.378$a_1^2$\\
$B_c^+\to B^+\bar K^0$ & 34$a_2^2$ & 286$a_2^2$ & 39.1$a_2^2$ &
96.5$a_2^2$ & 24.0$a_2^2$  & & 103.4$a_2^2$\\
$B_c^+\to B^+\bar K^{*0}$ & 13$a_2^2$ & 64$a_2^2$ & 46.8$a_2^2$ &
68.2$a_2^2$ & 13.8$a_2^2$  &  & 36.6$a_2^2$\\
$B_c^+\to B^{*+}\bar K^0$ & 15$a_2^2$ & 231$a_2^2$ & 24.0$a_2^2$ &
73.3$a_2^2$ & 8.9$a_2^2$  &  & 28.9$a_2^2$\\
$B_c^+\to B^{*+}\bar K^{*0}$ & 120$a_2^2$ & 242$a_2^2$ & 247$a_2^2$ &
141$a_2^2$ & 82.3$a_2^2$  & & 143.6$a_2^2$\\
$B_c^+\to B^+\pi^0$ & 0.5$a_2^2$ & 5.3$a_2^2$ & 0.51$a_2^2$ &
1.65$a_2^2$ & 1.03$a_2^2$ & & \\
$B_c^+\to B^+\rho^0$ & 0.7$a_2^2$ & 4.4$a_2^2$ & 1.40$a_2^2$ &
2.98$a_2^2$ & 1.28$a_2^2$ & & \\
$B_c^+\to B^{*+}\pi^0$ & 0.13$a_2^2$ & 4.8$a_2^2$ & 0.38$a_2^2$ &
1.45$a_2^2$ & 0.53$a_2^2$ & & \\
$B_c^+\to B^{*+}\rho^0$ & 3.4$a_2^2$ & 13.1$a_2^2$ & 4.50$a_2^2$ &
5.96$a_2^2$ & 4.56$a_2^2$ & &
\end{tabular}
\end{ruledtabular}
\end{table}

The matrix elements of the weak current between the $B_c$ meson and
the final $B_s^{(*)}$, $B$ meson entering the factorized nonleptonic decay
amplitude (\ref{factor}) are parametrized by the set of decay form
factors defined in Eqs.~(\ref{eq:pff1}) and (\ref{eq:vff1}). Using the
form factor values obtained in Sec.~\ref{dff}, we get predictions for
the nonleptonic $B_c^+\to F^0M^+$ and $B_c^+\to B^+M^0$  decay rates
and give them in Table~\ref{ssdr} in 
comparison with other calculations \cite{klo,emv,cc,aknt,cdf,lc}.

In Tables~\ref{sdr},~\ref{ssdr} we confront the predictions of our
model for semileptonic and nonleptonic $B_c$ decays with previous
calculations \cite{iks,klo,emv,cc,aknt,nw,cdf,lyl,lc}. The constituent
quark models of Refs.~\cite{iks,nw} are based on the same effective
quark-meson Lagrangian but use different phenomenological
parameterizations (Gaussian \cite{iks} and dipole \cite{nw}) for the
vertex functions, which are assumed to depend only on the loop
momentum flowing through the vertex. The relativistic quark models of
Refs.~\cite{emv,cc,lc} use different reductions of the
Bethe-Salpeter (BS) equation. The authors of Ref.~\cite{cc} apply a
nonrelativistic instantaneous approximation to the decay matrix
elements and relate the BS wave functions to the
Schr\"odinger ones. They do not give sufficient information about
the quark interaction potential in their model. The quark models
\cite{emv,lc} are based on the instantaneous approximation and
different versions of the quasipotential equation. The meson wave
functions are obtained by solving these equations with the one-gluon
exchange plus the long-range scalar linear potentials. The light-front
relativistic quark model is used in Ref.~\cite{aknt}. The decay form
factors are expressed through the overlap integrals of the meson
light-front wave functions which are related to the equal-time wave
functions. The latter are expressed through the Gaussian functions.
The heavy quark spin symmetry relations \cite{jlms} and constituent
quark models are used in Refs.~\cite{cdf,lyl}. This symmetry permits
to relate the $B_c$ weak decay form factors to a few invariant
functions near the zero recoil point. These invariant functions are
determined from the wave equation with the Richardson potential
\cite{cdf} or by the Gaussian wave functions \cite{lyl}. Then they are
extrapolated to the whole kinematical range accessible in $B_c$ decays.  
The authors of Ref.~\cite{klo} employ three-point QCD sum rules with
the account for the Coulomb-like $\alpha_s/v$-corrections. The values
of the form factors are determined in the vicinity of $q^2=0$ and then
are extrapolated to the allowed kinematical region using the pole
ansatz.

Our relativistic quark model provides the selfconsistent dynamical
approach  for the calculation of various meson properties. The meson
wave functions in this approach are obtained as the solutions of the
relativistic quasipotential equation. The weak decay matrix elements
are expressed in terms of these wave functions. It allows us to
determine explicitly the $q^2$ dependence of the form factors of the
weak $B_c$ decays in the whole kinematical range. All relevant
relativistic effects are taken into account (including the boost of
the wave functions to the moving reference frame). The light quarks in
the final heavy-light meson are treated relativistically. All that
increases the reliability of the obtained results.     

As one sees from Tables~\ref{sdr},~\ref{ssdr} the theoretical
predictions for $B_c$ weak decay rates differ substantially. Thus
experimental measurements of corresponding decay rates can
discriminate between various approaches.     

\section{Conclusions}
\label{sec:conc}

In this paper we considered weak semileptonic and nonleptonic $B_c$
decays to $B_s$ and $B$ mesons, associated with $c\to s,d$ quark
transition, in the framework of the relativistic
quark model based on the quasipotential approach in quantum field
theory. The weak decay form factors were calculated explicitly in the
whole kinematical range using the heavy quark expansion for the 
initial active quark $c$ and spectator quark $\bar b$. The final
quark $s$ or $d$ was treated completely relativistically without
applying an unjustified expansion in inverse powers of its mass. 
The leading order contribution of the heavy quark expansion was treated
exactly, while in calculating the subleading order contribution the
replacement of light quark energies $\epsilon_q(p)$ ($q=s,d$) by the
center of mass energies $E_q$ on mass shell  was performed. It was shown
that such substitution introduces only minor errors which are of
the same order as the higher order terms in the heavy quark expansion. Thus
the decay form factors were evaluated up to the subleading order of the
heavy quark expansion. The overall subleading contributions are small
and weakly depend on the momentum transfer $q^2$.

\begin{table}
\caption{Branching fractions (in \%) of exclusive $B_c$ decays calculated
  for the fixed values of the $B_c$ lifetime $\tau_{B_c}=0.46$~ps and
  $a_1=1.20$, $a_2=-0.317$.}
\label{Br}
\begin{ruledtabular}
\begin{tabular}{cccccc}
Decay& Br & Decay & Br & Decay & Br\\
\hline
$B_c\to B_s e\nu$& 0.84 &$B_c^+\to B_s K^{*+} $& 0.003  &
$B_c^+\to B^{*0} K^{*+}$& 0.033 \\ 
$B_c\to B^*_s e\nu$& 1.75 &$B_c^+\to B_s^* K^+ $& 0.11 & $B_c^+\to
B^+ \bar K^{0}$& 0.24 \\
$B_c\to B e\nu$& 0.042 &$B_c^+\to B^0\pi^+ $& 0.10  &$B_c^+\to
B^+ \bar K^{*0}$& 0.09\\
 $B_c\to B^* e\nu$& 0.12 &$B_c^+\to B^0 \rho^+$& 0.13 &
 $B_c^+\to B^{*+} \bar K^{0}$& 0.11 \\ 
$B_c^+\to B_s\pi^+$& 2.52 & $B_c^+\to B^{*0} \pi^+$& 0.026 &$B_c^+\to
B^{*+} \bar K^{*0}$& 0.84\\
$B_c^+\to B_s\rho^+$& 1.41 &$B_c^+\to B^{*0} \rho^{+}$& 0.68&$B_c^+\to
B^+ \pi^{0}$& 0.004\\
$B_c^+\to B_s^*\pi^+$&1.61 & $B_c^+\to B^0 K^{+}$& 0.009 & $B_c^+\to
B^+ \rho^{0}$& 0.005\\
$B_c^+\to B_s^*\rho^+$&11.1 &$B_c^+\to B^0 K^{*+}$& 0.004  &$B_c^+\to
B^{*+} \pi^{0}$& 0.001\\
$B_c^+\to B_s K^+$&0.21 &$B_c^+\to B^{*0} K^{+}$& 0.004  &$B_c^+\to
B^{*+} \rho^{0}$& 0.024\\
\end{tabular}
\end{ruledtabular}
\end{table}

We calculated semileptonic and nonleptonic (in factorization
approximation) $B_c$ decay rates. 
Our predictions for the branching fractions are summarized in
Table~\ref{Br}, where we use the central experimental value of the
$B_c$ meson lifetime \cite{pdg}. From this table we see that
the considered semileptonic 
decays to $B_s$ and $B$ mesons give in total $\sim2.0$\% of the $B_c$
decay rate, while the energetic nonleptonic decays provide
the dominant contribution $\sim19.3$\%. In our recent paper \cite{bcbd} we
calculated weak $B_c$ decays to charmonium and $D$ mesons, associated
with $\bar b\to \bar c, \bar u$ quark transition. It was found that
the semileptonic decays to the ground and first radially excited states
of charmonium and to $D$ mesons yield $\sim1.7$\% and corresponding
energetic nonleptonic decays (to charmonium and $K^{(*)}$ or
$\pi,\rho$ mesons) contribute $\sim0.6$\%. All these decays (to $B_s$, $B$,
charmonium and $D$ mesons) 
add up to $\sim 23.6$\% of the $B_c$ total decay rate.

\acknowledgments
The authors express their gratitude to M. M\"uller-Preussker and
V. Savrin for support and discussions.
Two of us (R.N.F and V.O.G.) were supported in part by the 
{\it Deutsche Forschungsgemeinschaft} under contract Eb 139/2-2.

\appendix*
\section{Form factors of weak $\bm{B_{\lowercase{c}}}$ decays}

(a) $B_c\to P$ transition ($P=B_s,B$)

\begin{eqnarray}
  \label{eq:fpl}
  f_+^{(1)}(q^2)&=&\sqrt{\frac{E_P}{M_{B_c}}}\int \frac{{\rm
  d}^3p}{(2\pi)^3} \bar\Psi_P\left({\bf p}+\frac{2m_b}{E_P+M_P}{\bf
  \Delta} \right)\sqrt{\frac{\epsilon_q(p+\Delta)+
  m_q}{2\epsilon_q(p+\Delta)}} \sqrt{\frac{\epsilon_c(p)+
  m_c}{2\epsilon_c(p)}}\cr
&&\times\Biggl\{1+\frac{M_{B_c}-E_P}{\epsilon_q(p+\Delta)+m_q}+\frac{({\bf
p\Delta})}{{\bf \Delta}^2}\Bigglb(\frac{{\bf
\Delta}^2}{[\epsilon_q(p+\Delta)+m_q][\epsilon_c(p)+m_c]}+(M_{B_c}-E_P) \cr
&&\times\left(\frac1{\epsilon_q(p+\Delta)+m_q}+
\frac1{\epsilon_c(p)+m_c}\right) \Biggrb)+\frac{{\bf
  p}^2}{[\epsilon_q(p+\Delta)+m_q][\epsilon_c(p)+m_c]} \cr
&&+\frac23{\bf p}^2\Bigglb(\frac{E_P-M_P}
{[\epsilon_q(p+\Delta)+m_q][\epsilon_c(p)+m_c]}
\left(\frac1{\epsilon_q(p+\Delta)+m_q}-\frac1{\epsilon_b(p)+m_b}\right)\cr
&&+\frac{M_{B_c}-E_P}{E_P+M_P}\left(\frac1{\epsilon_q(p+\Delta)+m_q}-
\frac1{\epsilon_c(p)+m_c}\right)\cr
&&\times\left(\frac1{\epsilon_b(p)+m_b}-
\frac1{\epsilon_q(p+\Delta)+m_q}\right)\Biggrb)\Biggr\}\Psi_{B_c}({\bf p}),  
\end{eqnarray}

\begin{eqnarray}
  \label{eq:fpls}
  f_+^{S(2)}(q^2)&=&\sqrt{\frac{E_P}{M_{B_c}}}\int \frac{{\rm
  d}^3p}{(2\pi)^3} \bar\Psi_P\left({\bf p}+\frac{2m_b}{E_P+M_P}{\bf
  \Delta} \right)\sqrt{\frac{E_q+
  m_q}{2E_q}} \Biggl\{
  \frac1{E_q}\Biggl[ \frac{E_q-m_q}
  {E_q+m_q}\left(1-\frac{E_q+m_q}{2m_c}\right) \cr
&& -\frac{M_{B_c}-E_P}{E_q+mq}\Biggr]
\left[M_P-\epsilon_q\left(p+\frac{2m_b}{E_P+M_P}\Delta\right)
-\epsilon_b\left(p+\frac{2m_b}{E_P+M_P}\Delta \right)\right]
  \cr
&&+\frac{({\bf p\Delta})}{{\bf \Delta}^2}\Bigglb(
\frac1{2E_q}\Biggl[\frac{{\bf \Delta}^2}
{(E_q+m_q)^2}
  -\frac{M_{B_c}-E_P}{E_q+m_q}\left(1
  +\frac{E_q-m_q} 
{2m_c}\right)\Biggr]
\Biggl[M_{B_c}+M_P-\epsilon_b(p)\cr
&&-\epsilon_c(p)
-\epsilon_q\left(p+\frac{2m_b}{E_P+M_P}\Delta\right)
-\epsilon_b\left(p+\frac{2m_b}{E_P+M_P}\Delta \right)\Biggr]
-\frac{{\bf \Delta}^2}{2m_cE_q(E_q+m_q)}\cr
&&\times
  \Biggl[M_P-\epsilon_q\left(p+\frac{2m_b}{E_P+M_P}\Delta\right)
-\epsilon_b\left(p+\frac{2m_b}{E_P+M_P}\Delta \right)\Biggr]
\Biggrb)\Biggr\}\Psi_{B_c}({\bf p}),
\end{eqnarray}

\begin{eqnarray}
  \label{eq:fplv}
  f_+^{V(2)}(q^2)&=&\sqrt{\frac{E_P}{M_{B_c}}}\int \frac{{\rm
  d}^3p}{(2\pi)^3} \bar\Psi_P\left({\bf p}+\frac{2m_b}{E_P+M_P}{\bf
  \Delta} \right)\sqrt{\frac{E_q+m_q}{2E_q}}\cr&&\times
\Biggl\{-\frac{(E_q-m_q)(m_Q+M_{B_c}-E_P)}{2m_bE_q(E_q+m_q)}
\Biggl[M_{B_c}+M_P-\epsilon_b(p)-\epsilon_c(p)\cr
&&
-\epsilon_q\left(p+\frac{2m_b}{E_P+M_P}\Delta\right)
-\epsilon_b\left(p+\frac{2m_b}{E_P+M_P}\Delta \right)\Biggr]
+\frac{({\bf p\Delta})}{{\bf
  \Delta}^2} \Bigglb(\frac{E_q-m_q}{2E_qm_q} \cr
&&\times
\Biggl(\frac{E_q-m_q}{E_q+m_q}\left[1-\frac{E_q+m_q}{2m_c}\right]
\Biggl[M_{B_c}-\epsilon_b(p)-\epsilon_c(p)\Biggr]
-\left[1-\frac{E_q-m_q}{2m_c}\right]\cr&&\times
\Biggl[M_P
-\epsilon_q\left(p+\frac{2m_b}{E_P+M_P}\Delta\right)
-\epsilon_b\left(p+\frac{2m_b}{E_P+M_P}\Delta \right)\Biggr]\Biggl)
-\frac{1}{2m_b}\Biggl[\frac{m_q}{E_q}\Biggl(\frac{E_q-m_q}{E_q+m_q}\cr
&&-
\frac{M_{B_c}-E_P}{E_q+m_q}\Biggr)+\frac{E_q-m_q}{E_q}
  \frac{M_{B_c}-E_P}{E_q+m_q}\Biggr]
\Biggl[M_{B_c}+M_P-\epsilon_b(p)-\epsilon_c(p)\cr
&&
-\epsilon_q\left(p+\frac{2m_b}{E_P+M_P}\Delta\right)
-\epsilon_b\left(p+\frac{2m_b}{E_P+M_P}\Delta \right)\Biggr]
\Biggrb)\Biggr\}\Psi_{B_c}({\bf p}),
\end{eqnarray}

\begin{eqnarray}
  \label{eq:f01}
  f_0^{(1)}(q^2)&=&\frac{2\sqrt{E_PM_{B_c}}}{M_{B_c}^2-M_P^2}\int \frac{{\rm
  d}^3p}{(2\pi)^3} \bar\Psi_P\left({\bf p}+\frac{2m_b}{E_P+M_P}{\bf
  \Delta} \right)\sqrt{\frac{\epsilon_q(p+\Delta)+
  m_q}{2\epsilon_q(p+\Delta)}} \sqrt{\frac{\epsilon_c(p)+
  m_c}{2\epsilon_c(p)}}\cr
&&\times\Biggl\{(E_P+M_P)\left(\frac{E_P-M_P}{\epsilon_q(p+\Delta)+m_q}+
\frac{M_{B_c}-E_P}{E_P+M_P}\right)+({\bf
p\Delta})\Bigglb(\frac{1}{\epsilon_q(p+\Delta)+m_q}\cr
&&+\frac{1}{\epsilon_c(p)+m_c}+\frac{M_{B_c}-E_P}
{[\epsilon_q(p+\Delta)+m_q][\epsilon_c(p)+m_c]} \Biggrb)+\frac{{\bf
  p}^2(M_{B_c}-E_P)}{[\epsilon_q(p+\Delta)+m_q][\epsilon_c(p)+m_c]} \cr
&&-\frac23{\bf p}^2(E_P-M_P)
\left(\frac1{\epsilon_q(p+\Delta)+m_q}-\frac1{\epsilon_b(p)+m_b}\right)
\Biggl(\frac1{\epsilon_q(p+\Delta)+m_q}\cr
&&-
\frac1{\epsilon_c(p)+m_c}-\frac{M_{B_c}-E_P}
{[\epsilon_q(p+\Delta)+m_q][\epsilon_c(p)+m_c]}\Biggr)
\Biggr\}\Psi_{B_c}({\bf p}),  
\end{eqnarray}

\begin{eqnarray}
\label{eq:f0s}
f_0^{S(2)}(q^2)&=&\frac{2\sqrt{E_PM_{B_c}}}{M_{B_c}^2-M_P^2}\int \frac{{\rm
d}^3p}{(2\pi)^3} \bar\Psi_P\left({\bf p}+\frac{2m_b}{E_P+M_P}{\bf
\Delta} \right)\sqrt{\frac{E_q+m_q}{2E_q}}\frac1{E_q(E_q+m_q)}\cr
&&\times\Biggl\{\left[(E_q-m_q)(M_{B_c}-E_P)\left(1
-\frac{E_q+m_q}{2m_c}\right)-{\bf\Delta}^2\right]
\Biggl[M_P-\epsilon_q\left(p+\frac{2m_b}{E_P+M_P}\Delta\right)\cr
&&
-\epsilon_b\left(p+\frac{2m_b}{E_P+M_P}\Delta \right)\Biggr]
  +\frac{({\bf p\Delta})}{2}\Bigglb(
\Biggl[\frac{M_{B_c}-E_P}{E_q+m_q}-1
-\frac{E_q-m_q}{2m_c}\Biggr]\cr 
&&\times\Biggl[M_{B_c}+M_P-\epsilon_b(p)-\epsilon_c(p)
-\epsilon_q\left(p+\frac{2m_b}{E_P+M_P}\Delta\right)
-\epsilon_b\left(p+\frac{2m_b}{E_P+M_P}\Delta \right)\Biggr]\cr
&&-\frac{1}{m_c}
  \left[M_{B_c}-E_P-2(E_q-m_q)\right] 
\Biggl[M_P-\epsilon_q\left(p+\frac{2m_b}{E_P+M_P}\Delta\right)
\cr &&
-\epsilon_b\left(p+\frac{2m_b}{E_P+M_P}\Delta \right)\Biggr]
\Biggrb)\Biggr\}\Psi_{B_c}({\bf p}), 
\end{eqnarray}

\begin{eqnarray}
\label{eq:f0v}
f_0^{V(2)}(q^2)&=&\frac{2\sqrt{E_PM_{B_c}}}{M_{B_c}^2-M_P^2}\int \frac{{\rm
d}^3p}{(2\pi)^3} \bar\Psi_P\left({\bf p}+\frac{2m_b}{E_P+M_P}{\bf
\Delta} \right)\sqrt{\frac{E_q+m_q}{2E_q}}\frac1{2E_q(E_q+m_q)}\cr
&&
\times\Biggl\{-\frac{E_q-m_q}{m_b}[m_q(M_{B_c}-E_{B_c})+{\bf\Delta}^2]
\Biggl[M_{B_c}+M_P-\epsilon_b(p)-\epsilon_c(p)\cr
&&-\epsilon_q\left(p+\frac{2m_b}{E_P+M_P}\Delta\right)
-\epsilon_b\left(p+\frac{2m_b}{E_P+M_P}\Delta \right)\Biggr]
+({\bf p\Delta})\Bigglb(\frac{M_{B_c}-E_{B_c}}{m_q}\cr
&&\times \Biggl(\frac{E_q-m_q}{E_q+m_q}
\left[1-\frac{E_q+m_q}{2m_c}\right]
[M_{B_c}-\epsilon_b(p)-\epsilon_c(p)]
- \left[1-\frac{E_q-m_q}{2m_c}\right]\cr
&&\times
\Biggl[M_P-\epsilon_q\left(p+\frac{2m_b}{E_P+M_P}\Delta\right)
-\epsilon_b\left(p+\frac{2m_b}{E_P+M_P}\Delta \right)\Biggr]\Biggr)
-\frac1{m_b}\Biggl[\frac{m_q}{E_q+m_q}\cr
&&\times
(M_{B_c}-E_{B_c}-E_q-m_q)+E_q-m_q\Biggr]
\Biggl[M_{B_c}+M_P-\epsilon_b(p)-\epsilon_c(p)\cr
&&-\epsilon_q\left(p+\frac{2m_b}{E_P+M_P}\Delta\right)
-\epsilon_b\left(p+\frac{2m_b}{E_P+M_P}\Delta \right)\Biggr] 
\Biggrb)
\Biggr\}\Psi_{B_c}({\bf p}),
\end{eqnarray}
where 
\[ \left|{\bf \Delta}\right|=\sqrt{\frac{(M_{B_c}^2+M_P^2-q^2)^2}
{4M_{B_c}^2}-M_P^2},\]
\[ E_P=\sqrt{M_P^2+{\bf \Delta}^2},\quad \epsilon_Q(p+\lambda
\Delta)=\sqrt{m_Q^2+({\bf p}+\lambda{\bf \Delta})^2} \quad (Q=b,s,d), \]
and the subscript $q$ corresponds to $s$ or $d$ quark for the final
$B_s$ or $B$ meson, respectively. 

\bigskip
(b) $B_c\to V$ transition ($V=B_s^*,B^*$)

\begin{eqnarray}
  \label{eq:v1}
  V^{(1)}(q^2)&=&\frac{M_{B_c}+M_V}{2\sqrt{M_{B_c}M_V}}\int \frac{{\rm
  d}^3p}{(2\pi)^3} \bar\Psi_V\left({\bf p}+\frac{2m_b}{E_V+M_V}{\bf
  \Delta} \right)\sqrt{\frac{\epsilon_q(p+\Delta)+
  m_q}{2\epsilon_q(p+\Delta)}} \sqrt{\frac{\epsilon_c(p)+
  m_c}{2\epsilon_c(p)}}\cr
&&\times\frac{2\sqrt{E_VM_V}}{\epsilon_q(p+\Delta)+m_q}\Biggl\{1
+\frac{({\bf p\Delta})}{{\bf\Delta}^2}\left(1-\frac{\epsilon_q(p+
\Delta)+m_q}{2m_c}\right)+\frac23\frac{{\bf p}^2}{E_V+M_V}\cr
&&\times\Biggl(
\frac{\epsilon_q(p+\Delta)+m_q}{2m_c[\epsilon_b(p)+m_b]}
-\frac1{\epsilon_q(p+\Delta)+m_q}\Biggr)\Biggr\}\Psi_{B_c}({\bf p}),
\end{eqnarray}

\begin{eqnarray}
  \label{eq:vs}
  V^{S(2)}(q^2)&=&\frac{M_{B_c}+M_V}{2\sqrt{M_{B_c}M_V}}\int \frac{{\rm
  d}^3p}{(2\pi)^3} \bar\Psi_V\left({\bf p}+\frac{2m_b}{E_V+M_V}{\bf
  \Delta} \right)\sqrt{\frac{E_q+m_q}{2E_q}}\frac{2\sqrt{E_VM_V}}
{E_q+m_q}\cr
&&\times\Biggl\{-\frac1{E_q}\left(1+\frac{E_q-m_q}{4m_c}\right)
\left[M_V-\epsilon_q\left(p+\frac{2m_b}{E_V+M_V}\Delta\right)
-\epsilon_b\left(p+\frac{2m_b}{E_V+M_V}\Delta \right)\right]
 \cr
&&-\frac{({\bf p\Delta})}{{\bf \Delta}^2}
\Bigglb(\frac{1}{2E_q}
\Biggl[M_{B_c}+M_V-\epsilon_b(p)-\epsilon_c(p)
-\epsilon_q\left(p+\frac{2m_b}{E_V+M_V}\Delta\right)\cr
&&
-\epsilon_b\left(p+\frac{2m_b}{E_V+M_V}\Delta \right)\Biggr]
+\frac{E_q-m_q}{2m_cE_q}
\Biggl[M_V-\epsilon_q\left(p+\frac{2m_b}{E_V+M_V}\Delta\right)\cr 
&&
-\epsilon_b\left(p+\frac{2m_b}{E_V+M_V}\Delta \right)\Biggr]
\Biggrb) \Biggr\}\Psi_{B_c}({\bf p}), 
\end{eqnarray}

\begin{eqnarray}
  \label{eq:vv}
V^{V(2)}(q^2)&=&\frac{M_{B_c}+M_V}{2\sqrt{M_{B_c}M_V}}\int \frac{{\rm
d}^3p}{(2\pi)^3} \bar\Psi_V\left({\bf p}+\frac{2m_b}{E_V+M_V}{\bf
\Delta} \right)\sqrt{\frac{E_q+m_q}{2E_q}}\frac{2\sqrt{E_VM_V}}
{E_q+m_q}\cr
&&\times\Biggl\{\frac{E_q-m_q}{4E_qm_c}
\left[M_V-\epsilon_q\left(p+\frac{2m_b}{E_V+M_V}\Delta\right)
-\epsilon_b\left(p+\frac{2m_b}{E_V+M_V}\Delta \right)\right]\cr
&&
-\frac{({\bf p\Delta})}{{\bf \Delta}^2}\Bigglb(\frac{E_q-m_q}{4E_qm_q}
\left(1+\frac{E_q-m_q}{2m_c}\right)
\Biggl[M_{B_c}-M_V-\epsilon_b(p)-\epsilon_c(p)\cr
&&
+\epsilon_q\left(p+\frac{2m_b}{E_V+M_V}\Delta\right)
+\epsilon_b\left(p+\frac{2m_b}{E_V+M_V}\Delta \right)\Biggr]
-\frac{E_q+m_q}{4E_qm_b}\Biggl[M_{B_c}+M_V
-\epsilon_b(p)\cr
&&-\epsilon_c(p)-
\epsilon_q\left(p+\frac{2m_b}{E_V+M_V}\Delta\right)
-\epsilon_b\left(p+\frac{2m_b}{E_V+M_V}\Delta \right)\Biggr]\Biggrb)
\Biggr\}\Psi_{B_c}({\bf p}), 
\end{eqnarray}

\begin{eqnarray}
  \label{eq:a11}
A_1^{(1)}(q^2)&=&\frac{2\sqrt{M_{B_c}M_V}}{M_{B_c}+M_V}
\sqrt{\frac{E_V}{M_V}}\int \frac{{\rm
d}^3p}{(2\pi)^3} \bar\Psi_V\left({\bf p}+\frac{2m_b}{E_V+M_V}{\bf
\Delta} \right)\sqrt{\frac{\epsilon_q(p+\Delta)+
m_q}{2\epsilon_q(p+\Delta)}} \sqrt{\frac{\epsilon_c(p)+
m_c}{2\epsilon_c(p)}}\cr
&&\times\Biggl\{1+\frac{1}{2m_c[\epsilon_q(p+\Delta)+m_q]}
\left[\frac23{\bf p}^2\frac{E_V-M_V}{\epsilon_b(p)+m_b}-
\frac{{\bf p}^2}3-
({\bf p\Delta})\right]\Biggr\}\Psi_{B_c}({\bf p}),
\end{eqnarray}

\begin{eqnarray}
  \label{eq:a1s}
  A_1^{S(2)}(q^2)&=&\frac{2\sqrt{M_{B_c}M_V}}{M_{B_c}+M_V}
\sqrt{\frac{E_V}{M_V}}\int \frac{{\rm
d}^3p}{(2\pi)^3} \bar\Psi_V\left({\bf p}+\frac{2m_b}{E_V+M_V}{\bf
\Delta} \right)\sqrt{\frac{E_q+m_q}{2E_q}}
\cr &&\times\frac{E_q-m_q}{E_q(E_q+m_q)}\left(1
+\frac{E_q-m_q}{2m_c}\right)
\Biggl[M_V-\epsilon_q\left(p+\frac{2m_b}{E_V+M_V}\Delta\right)\cr
&&
-\epsilon_b\left(p+\frac{2m_b}{E_V+M_V}\Delta \right)\Biggr]
\Psi_{B_c}({\bf p}), 
\end{eqnarray}

\begin{eqnarray}
  \label{eq:a1v}
  A_1^{V(2)}(q^2)&=&\frac{2\sqrt{M_{B_c}M_V}}{M_{B_c}+M_V}
\sqrt{\frac{E_V}{M_V}}\int \frac{{\rm
d}^3p}{(2\pi)^3} \bar\Psi_V\left({\bf p}+\frac{2m_b}{E_V+M_V}{\bf
\Delta} \right)\sqrt{\frac{E_q+m_q}{2E_q}}
\ \frac{E_q-m_q}{2E_q(E_q+m_q)}\cr 
&&\times\frac{({\bf p\Delta})}{{\bf \Delta}^2}
\Biggl\{-\left(1+\frac{m_q}{m_b}\right)\Biggl[M_{B_c}+M_V-
\epsilon_b(p)-\epsilon_c(p)
-\epsilon_q\left(p+\frac{2m_b}{E_V+M_V}\Delta\right)\cr
&&
-\epsilon_b\left(p+\frac{2m_b}{E_V+M_V}\Delta \right)\Biggr]
+\frac{E_q}{m_q}\left(1+\frac{E_q^2-m_q^2}{2E_qm_c}\right)
\Biggl[M_{B_c}-M_V
-\epsilon_b(p)\cr &&-\epsilon_c(p)
+\epsilon_q\left(p+\frac{2m_b}{E_V+M_V}\Delta\right)
+\epsilon_b\left(p+\frac{2m_b}{E_V+M_V}\Delta \right)\Biggr]
\Biggr\}\Psi_{B_c}({\bf p}), 
\end{eqnarray}

\begin{eqnarray}
  \label{eq:a21}
  A_2^{(1)}(q^2)&=&\frac{M_{B_c}+M_V}{2\sqrt{M_{B_c}M_V}}
\frac{2\sqrt{E_VM_V}}{E_V+M_V}\int \frac{{\rm
  d}^3p}{(2\pi)^3} \bar\Psi_V\left({\bf p}+\frac{2m_b}{E_V+M_V}{\bf
  \Delta} \right)\sqrt{\frac{\epsilon_q(p+\Delta)+
  m_q}{2\epsilon_q(p+\Delta)}}\cr
&&\times \sqrt{\frac{\epsilon_c(p)+ m_c}{2\epsilon_c(p)}}
\Biggl\{1+\frac{M_V}{M_{B_c}}\left(1-\frac{E_V+M_V}{\epsilon_q(p+
\Delta)+m_q}\right)
-\frac{({\bf p\Delta})}{{\bf\Delta}^2}\frac{E_V+M_V}{\epsilon_q(p+
\Delta)+m_q}\cr
&&\times\Bigglb(\frac{E_V+M_V}{2m_c}
\left[1-\frac{M_V}{M_{B_c}}\left(1-\frac{\epsilon_q(p+
\Delta)+m_q}{E_V+M_V}\right)\right]+\frac{M_V}{M_{B_c}}\Biggrb)
\cr
&&+\frac23\frac{{\bf p}^2}{\epsilon_q(p+\Delta)+m_q}
\Bigglb(\frac1{2m_c}\Biggl[\frac{E_V+M_V}{\epsilon_b(p)+m_b}
-\frac12+\frac{M_V}{\epsilon_q(p+\Delta)+m_q}\cr
&&+
\frac{M_V}{M_{B_c}}\left(
\frac{\epsilon_q(p+\Delta)+m_q}{\epsilon_b(p)+m_b}-\frac{E_V+M_V}
{\epsilon_b(p)+m_b}+\frac12
-\frac{E_V}{\epsilon_q(p+\Delta)+m_q}\right)\Biggr]\cr
&&+\frac{M_V}{M_{B_c}}\left(\frac1{\epsilon_q(p+\Delta)+m_q}
+\frac1{\epsilon_b(p)+m_b}\right)\Biggrb)\Biggr\}\Psi_{B_c}({\bf p}),
\end{eqnarray}

\begin{eqnarray}
  \label{eq:a2s}
  A_2^{S(2)}(q^2)&=&\frac{M_{B_c}+M_V}{2\sqrt{M_{B_c}M_V}}
\frac{2\sqrt{E_VM_V}}{E_V+M_V}\int \frac{{\rm
  d}^3p}{(2\pi)^3} \bar\Psi_V\left({\bf p}+\frac{2m_b}{E_V+M_V}{\bf
\Delta} \right)\sqrt{\frac{E_q+m_q}{2E_q}}\cr
&&\times\Biggl\{\frac{E_q-m_q}{E_q(E_q+m_q)}
\Biggl[1+\frac{E_q-m_q}{2m_c}+\frac{M_V}{M_{B_c}}
\left(1+\frac{E_V+M_V}{E_q-m_q}+\frac{E_q-m_q}{2m_c}\right)\Biggr]\cr
&&\times
\left[M_V-\epsilon_q\left(p+\frac{2m_b}{E_V+M_V}\Delta\right)
-\epsilon_b\left(p+\frac{2m_b}{E_V+M_V}\Delta \right)\right]\cr
&& -\frac{({\bf p\Delta})}{{\bf \Delta}^2}\frac{M_V}{M_{B_c}}
\frac{E_V+M_V}{E_q(E_q+m_q)}
\Bigglb(\frac12
\left(1-\frac{E_q-m_q}{2m_c}\right)
\Biggl[M_{B_c}+M_V-\epsilon_b(p)-\epsilon_c(p)\cr
&&
-\epsilon_q\left(p+\frac{2m_b}{E_V+M_V}\Delta\right)
-\epsilon_b\left(p+\frac{2m_b}{E_V+M_V}\Delta \right)\Biggr]
-\frac{M_{B_c}-E_V}{m_c}
\Biggl[M_V\cr
&&-\epsilon_q\left(p+\frac{2m_b}{E_V+M_V}\Delta\right)
-\epsilon_b\left(p+\frac{2m_b}{E_V+M_V}\Delta \right)\Biggr]
\Biggrb)\Biggr\}\Psi_{B_c}({\bf p}), 
\end{eqnarray}

\begin{eqnarray}
  \label{eq:a2v}
  A_2^{V(2)}(q^2)&=&\frac{M_{B_c}+M_V}{2\sqrt{M_{B_c}M_V}}
\frac{2\sqrt{E_VM_V}}{E_V+M_V}\int \frac{{\rm
  d}^3p}{(2\pi)^3} \bar\Psi_V\left({\bf p}+\frac{2m_b}{E_V+M_V}{\bf
\Delta} \right)\sqrt{\frac{E_q+m_q}{2E_q}}\cr 
&&\times
\frac{({\bf p\Delta})}{{\bf \Delta}^2}\frac{E_q-m_q}{2E_q(E_q+m_q)}
\Biggl\{-\Bigglb(1
+\frac{m_q}{m_b}+\frac{M_V}{M_{B_c}}\left[1+\frac{m_q}{m_b}
\left(1+\frac{E_V+M_V}{E_q-m_q}\right)\right]\Biggrb)\cr
&&\times
\Biggl[M_{B_c}+M_V-\epsilon_b(p)-\epsilon_c(p)
-\epsilon_q\left(p+\frac{2m_b}{E_V+M_V}\Delta\right)
-\epsilon_b\left(p+\frac{2m_b}{E_V+M_V}\Delta \right)\Biggr]
\cr
&&+\frac{E_q}{m_q}\left[1+\frac{E_q^2-m_q^2}{2E_qm_q}
+\frac{M_V}{M_{B_c}}\left(1+
\frac{E_V+M_V}{E_q}+\frac{E_q^2-m_q^2}{2E_qm_q}\right)\right]
\Biggl[M_{B_c}-M_V-\epsilon_b(p)\cr
&&-\epsilon_c(p)
+\epsilon_q\left(p+\frac{2m_b}{E_V+M_V}\Delta\right)
+\epsilon_b\left(p+\frac{2m_b}{E_V+M_V}\Delta \right)\Biggr]\Biggrb)
\Biggr\}\Psi_{B_c}({\bf p}), 
\end{eqnarray}

\begin{eqnarray}
  \label{eq:a01}
  A_0^{(1)}(q^2)&=&\sqrt{\frac{E_V}{M_V}}\int \frac{{\rm
  d}^3p}{(2\pi)^3} \bar\Psi_V\left({\bf p}+\frac{2m_b}{E_V+M_V}{\bf
  \Delta} \right)\sqrt{\frac{\epsilon_q(p+\Delta)+
  m_q}{2\epsilon_q(p+\Delta)}} \sqrt{\frac{\epsilon_c(p)+
  m_c}{2\epsilon_c(p)}}\cr
&&\times\Biggl\{1+\frac{M_{B_c}-E_V}{\epsilon_q(p+\Delta)+m_q}
\Bigglb(1+\Biggl[\frac{({\bf p\Delta})}{{\bf \Delta}^2}
-\frac23\frac{{\bf p}^2}{E_V+M_V}
\Biggl(\frac{1}{\epsilon_q(p+\Delta)+m_q}\cr
&&
+\frac{1}{\epsilon_b(p)+m_b}\Biggr)\Biggr]
\left[1+\frac1{2m_c}\left(\frac{{\bf \Delta}^2}{M_{B_c}-E_V}
+\epsilon_q(p+\Delta)+m_q\right)\right]\Biggrb)\cr
&&-\frac{{\bf p}^2}{6m_c[\epsilon_q(p+\Delta)+m_q]}\Biggr\}
\Psi_{B_c}({\bf p}),
\end{eqnarray}

\begin{eqnarray}
  \label{eq:a0s}
A_0^{S(2)}(q^2)&=&\sqrt{\frac{E_V}{M_V}}\int \frac{{\rm
d}^3p}{(2\pi)^3} \bar\Psi_V\left({\bf p}+\frac{2m_b}{E_V+M_V}{\bf
\Delta} \right)\sqrt{\frac{E_q+m_q}{2E_q}}\frac1{E_q(E_q+m_q)}
\Biggl\{\Biggl[(E_q-m_q)\cr
&&\times\left(1-\frac{E_q-m_q}{2m_c}\right)-M_{B_c}+E_V\Biggr]
\Biggl[M_V-\epsilon_q\left(p+\frac{2m_b}{E_V+M_V}\Delta\right)\cr
&& 
-\epsilon_b\left(p+\frac{2m_b}{E_V+M_V}\Delta \right)\Biggl]
-\frac{({\bf p\Delta})}{{\bf \Delta}^2}
\Bigglb(\frac{M_{B_c}-E_V}2
\left(1-\frac{E_q-m_q}{2m_c}\right)
\Biggl[M_{B_c}+M_V\cr
&&-\epsilon_b(p)-\epsilon_c(p)
-\epsilon_q\left(p+\frac{2m_b}{E_V+M_V}\Delta\right)
-\epsilon_b\left(p+\frac{2m_b}{E_V+M_V}\Delta \right)\Biggr]\cr
&&
+\frac{{\bf\Delta}^2}{m_c} 
\Biggl[M_V-\epsilon_q\left(p+\frac{2m_b}{E_V+M_V}\Delta\right)
-\epsilon_b\left(p+\frac{2m_b}{E_V+M_V}\Delta \right)\Biggr]
\Biggrb)\Biggr\}\Psi_{B_c}({\bf p}), 
\end{eqnarray}

\begin{eqnarray}
  \label{eq:a0v}
A_0^{V(2)}(q^2)&=&\sqrt{\frac{E_V}{M_V}}\int \frac{{\rm
d}^3p}{(2\pi)^3} \bar\Psi_V\left({\bf p}+\frac{2m_b}{E_V+M_V}{\bf
\Delta} \right)\sqrt{\frac{E_q+m_q}{2E_q}}
\frac{({\bf p\Delta})}{{\bf \Delta}^2}\frac1{2E_q(E_q+m_q)}
\cr&&\times
\Biggl\{\frac1{m_q}
\left[\left(\frac{E_q-m_q}{2m_c}+\frac{E_q}{E_q+m_q}\right)
{\bf\Delta}^2-(E_q-m_q)(M_{B_c}-E_V)\right]
\Biggl[M_{B_c}-M_V\cr
&&-\epsilon_b(p)-\epsilon_c(p)
+\epsilon_q\left(p+\frac{2m_b}{E_V+M_V}\Delta\right)
+\epsilon_b\left(p+\frac{2m_b}{E_V+M_V}\Delta \right)\Biggr]\cr
&&
-\left[\frac{{\bf\Delta}^2}{E_q+m_q}+\frac{m_q}{m_b}
\left(\frac{{\bf\Delta}^2}{E_q+m_q}-M_{B_c}+E_V\right)\right]
\Biggl[M_{B_c}+M_V-\epsilon_b(p)-\epsilon_c(p)\cr
&&
-\epsilon_q\left(p+\frac{2m_b}{E_V+M_V}\Delta\right)
-\epsilon_b\left(p+\frac{2m_b}{E_V+M_V}\Delta \right)\Biggr]
\Biggr\}\Psi_{B_c}({\bf p}), 
\end{eqnarray}
where 
\[ \left|{\bf \Delta}\right|=\sqrt{\frac{(M_{B_c}^2+M_V^2-q^2)^2}
{4M_{B_c}^2}-M_V^2},\]
\[ E_V=\sqrt{M_V^2+{\bf \Delta}^2}. \]


\begin{thebibliography}{99}
\bibitem{cdfcol} CDF Collaboration, F. Abe {\it et al.}, Phys. Rev. D
  {\bf 58}, 112004 (1998). 
\bibitem{gklry} I. P. Gouz, V. V. Kiselev, A. K. Likhoded,
  V. I. Romanovsky, and O. P. Yushchenko, hep-ph/0211432.
\bibitem{bcbd} D. Ebert, R. N. Faustov and V. O. Galkin, hep-ph/0306306.
\bibitem{mass1} V. O. Galkin, A. Yu. Mishurov and R. N. Faustov, Yad. Fiz.
{\bf 55}, 2175 (1992) [Sov. J. Nucl. Phys. {\bf 55}, 1207 (1992)].
\bibitem{mass}  D. Ebert, R. N. Faustov and V. O. Galkin, Phys. Rev. D
  {\bf 62}, 034014 (2000).
\bibitem{hlm} D. Ebert, V. O. Galkin and R. N. Faustov, Phys. Rev. D
  {\bf 57}, 5663 (1998); {\bf 59}, 019902(E) (1999).
\bibitem{gf} V. O. Galkin and R. N. Faustov, Yad. Fiz. {\bf 44}, 1575
(1986) [Sov. J. Nucl. Phys. {\bf 44}, 1023 (1986)]; D. Ebert,
R. N. Faustov and V. O. Galkin, Phys. Lett. B {\bf 537}, 241 (2002).
\bibitem{fg} R. N. Faustov and V. O. Galkin, Z. Phys. C {\bf 66}, 119
(1995);  D. Ebert, R. N. Faustov and V.~O. Galkin, Phys. Rev. D {\bf
  62}, 014032 (2000).
\bibitem{mod} D. Ebert, R. N. Faustov and V. O. Galkin, Phys. Rev. D
  {\bf 56}, 312 (1997); R. N. Faustov, V.~O. Galkin and
  A. Yu. Mishurov, Phys. Rev. D {\bf 53}, 6302 (1996); {\bf  53}, 1391
  (1996).
\bibitem{efgbc}  D. Ebert, R. N. Faustov and V. O. Galkin, Phys. Rev. D
  {\bf 67}, 014027 (2003).
\bibitem{3} A. A. Logunov and A. N. Tavkhelidze, Nuovo Cimento {\bf29},
380 (1963).
\bibitem{4} A. P. Martynenko and R. N. Faustov, Theor. Math. Phys. {\bf
    64}, 765 (1985) [Teor. Mat. Fiz. {\bf 64}, 179 (1985)].
\bibitem{ef} E. Eichten and F. Feinberg, Phys. Rev. D {\bf 23},
2724 (1981).
\bibitem{schn} H. J. Schnitzer, Phys. Rev. D {\bf 18}, 3482 (1978).
\bibitem{f} R. N. Faustov, Ann. Phys. {\bf 78}, 176 (1973); Nuovo
Cimento A {\bf 69}, 37 (1970).
\bibitem{jlms} E. Jenkins, M. Luke, A. V. Manohar and M. Savage,
  Nucl. Phys. B {\bf 390}, 463 (1993).
\bibitem{pdg}  Particle Data Group, K. Hagiwara {\it et al.},
  Phys. Rev. D {\bf 66}, 010001 (2002).
\bibitem{iks} M. A. Ivanov, J. G. K\"orner and P. Santorelli,
  Phys. Rev. D {\bf 63}, 074010 (2001).
\bibitem{klo} V. V. Kiselev, A. E. Kovalsky and A. K. Likhoded,
  Nucl. Phys. B {\bf 585}, 353 (2000); V. V. Kiselev, hep-ph/0211021. 
\bibitem{emv} A. Abd El-Hady, J. H. Mu\~noz and J. P. Vary, Phys. Rev. D
  {\bf 62}, 014019 (2000).
\bibitem{cc} C.-H. Chang and Y.-Q. Chen, Phys. Rev. D {\bf 49}, 3399
  (1994). 
\bibitem{cdf} P. Colangelo and F. De Fazio, Phys. Rev. D {\bf 61},
  034012 (2000).
\bibitem{aknt} A. Yu. Anisimov, P. Yu. Kulikov, I. M. Narodetskii and
  K. A. Ter-Martirosyan, Phys. Atom. Nucl. {\bf 62}, 1739 (1999)
  [Yad. Fiz. {\bf 62}, 1868 (1999)].
\bibitem{nw} M. A. Nobes and R. M. Woloshyn, J. Phys. G {\bf 26}, 1079
  (2000). 
\bibitem{lyl} G. Lu, Y. Yang and H. Li, Phys. Lett. B {\bf 341}, 391
  (1995). 
\bibitem{lc} J.-F. Liu and K.-T. Chao, Phys. Rev. D {\bf 56}, 4133
  (1997). 
\bibitem{bsw} M. Bauer, B. Stech, and M. Wirbel, Z. Phys. C {\bf 34},
103 (1987).
\bibitem{dg} M. J. Dugan and B. Grinstein, Phys. Lett. B {\bf 255}, 583
(1991).
\bibitem{jb} J. D. Bjorken, Nucl. Phys. B (Proc. Suppl.) {\bf 11}, 325
(1989). 
\bibitem{bbns} M. Beneke, G. Buchalla, M. Neubert and C. T. Sachrajda,
Phys. Rev. Lett. {\bf 83}, 1914 (1999); Nucl. Phys. B {\bf 591}, 313
(2000). 
\bibitem{bs} A. J. Buras and L. Silvestrini, Nucl. Phys. B {\bf 569},
  3 (2000).

\end{thebibliography}
\end{document}